    \documentstyle[twocolumn,aps]{revtex}
\input myepsf.sty
\input myepsfig.sty
\def\tG{\tilde G}
\def\tF{\tilde F}

\def\v{\bf}
\def\ka{\kappa}
\def\ri{\right}
\def\le{\left}
\def\b1sum{\stackrel{\rule{.16in}{0.007in}}\sum}

\def\W{\Omega}
\def\w{\omega}
\def\t{\tau}
\def\si{\sigma}
\def\c{ \!\cdot\! }
\def\prod(#1,#2){ (#1\c #2) }
\def\ep{\epsilon}
\def\e{u}
\def\U{ {\cal P} }

\def\W{ {\cal W} }
\def\H{ {\cal H} }
\def\C{ {\cal C} }
\def\V{ {\cal V} }
\def\X{ {\cal X} }
\def\O{\Omega}
\def\tr{ {\rm Tr} }
\def\bk{\bar k}
\def\E{\bar E_k}
\def\md{ M_d }

\def\vk{{\v k}}
\def\k{|\vk|}
\def\bq{|{\bf q}|}
\def\x{ \xi } 
\def\m{ \mu } 
\def\Ek{E_k}
\def\En{\bar E_k}
\def\n{ \nu } 
\def\mn{ {\mu\nu} } 
\def\r{ \rho } 
\def\l{ \lambda } 
\def\rl{ {\rho\lambda} } 
\def\q{ r } 
\def\D{ {\cal D} }
\def\B{ {B\over m} }
\def\G{ {G\over m} }
\def\g{ \gamma }
\def\thru#1{\mathrel{\mathop{#1\!\!\!/\!}}}
\def\s#1{\thru{#1}}
\def\nnn{\nonumber}
\def\equ(#1){Eq.\ (\ref{#1})}
\def\bea{\begin{eqnarray}}
\def\eea{\end{eqnarray}}
\def\be{\begin{equation}}

\def\ee{\end{equation}}
\def\z(#1){ { \thru{#1} + m\over 2m} }
\def\zm(#1){ { \thru{#1} - m\over 2m} }
\def\kp{{k}_+}
\def\km{{k}_-}
\def\fp{f_+}
\def\fm{f_-}
\def\us{ u({\v k},s) }
\def\bus{ \bar u({\v k},s) }
\def\vs{ v(-{\v k},s) }
\def\bvs{ \bar v(-{\v k},s) }

\def\bur{ \bar u({\v k},r) }

\def\a2{\alpha^2}

\begin{document}

\title {\bf Parity-violating quasifree electron-deuteron scattering
            in a covariant approach.
       }  

  \author{ Grigorios I. Poulis
       } 
\address { 
        Department of Physics and Mathematical Physics\\
                and Special Research Center for the Subatomic
                 Structure of Matter\\
                University of Adelaide, 
                SA 5005, Australia\\
         }
\vskip 1 true cm
\bigskip
\date{Report \#: ADP-96-47/T241, to appear in {\it Few Body Systems} }

\maketitle       

\begin{abstract}{\em }

The role of relativistic corrections associated with 
lower Dirac components of the deuteron wavefunction,
is examined for parity-violating (PV) electron scattering.
The relation between these corrections 
and negative energy components of the struck nucleon
in relativistic PWIA is elucidated. The model dependence
induced by describing such effects, using different 
deuteron vertex functions and prescriptions for
the half-off-shell nucleon vertices, 
is compared against the precision required for studies of
nucleon strangeness in quasielastic PV $e-d$ scattering.

\end{abstract}

\section{Motivation}

Measuring the parity-violating (PV) longitudinal asymmetry 
in electron scattering offers the possibility to 
extract the hitherto unknown vector, $G^s_{E,M}$, 
and (though to much less extent) axial 
vector, $G^s_{A}$, strangeness nucleon form factors.
Several such measurements will be performed
in medium energy facilities, such as MIT-Bates~\cite{Bates}, 
Mainz-MAMI~\cite{MAMI} and TJNAF~\cite{TJNAF}. 
The first phase of the SAMPLE experiment~\cite{SAMPLE} at MIT-Bates
is currently under way and will attempt to extract $G_M^s$ 
from low momentum transfer, $q^2=-0.1$ (GeV/c)$^2$, backward angle,
elastic PV $e-p$ scattering. However, strong correlation with
the neutral current (NC), axial, isovector form 
factor, $\tilde G_A^{T=1}$, whose radiative 
corrections have a large theoretical uncertainty in
the Standard Model~\cite{rat1}, places constraints on the
effectiveness of the SAMPLE measurement. 
Hadjimichael et al. suggested~\cite{Hadj} 
that a complimentary measurement of the 
PV asymmetry in quasielastic scattering from deuterium,
may, in conjunction with the proton measurement, 
allow one to determine both $G^s_M$ and $\tilde G_A^{T=1}$.
The rationale of this suggestion is that, although 
both $G^s_M$ and $\tilde G_A^{T=1}$ appear in the
(predominantly transverse) backward asymmetry multiplied 
by magnetic ($G_M^{p,n}$) form factors, they come with different
weighting for a nuclear target, i.e., $(ZG_M^p+NG_M^n)$ {\it vs.} 
 $(ZG_M^p-NG_M^n)$, being, respectively, isoscalar and
isovector. Thus, for deuterium, this correlation is 
suppressed relatively to the proton by approximately 
$(\m_p -\m_n) /(\m_p + \m_n)> 5$. 
Following this analysis, SAMPLE (in its second phase) 
as well as the E91-017 experiment at
TJNAF~\cite{TJNAF}, and possibly a future 
experiment at Mainz~\cite{MAMI} 
will perform such measurements on deuterium. 

In attempting to extract such subtle subnuclear effects 
from a {\it nuclear} observable, it is imperative that the 
nuclear physics modeling can be trusted to the 1-2\% level.
In Ref.~\cite{Hadj} the theoretical uncertainty 
was estimated for the deuteron case, and was concluded 
that relativistic, rather than rescattering effects,
are the main source of model dependence in the PV asymmetry 
for quasifree kinematics at $\bq>0.7$ GeV/c, 
especially at forward angles ($6$\% effects). This follows from
a comparison of a traditional potential calculation including
final-state interactions (FSI), where
the current operators were truncated to first nontrivial
order in a nonrelativistic expansion, with a plane wave impulse
approximation  (PWIA) calculation, in which a nonrelativistic
momentum distribution is convoluted with a relativistic off-shell
single nucleon tensor obtained by a generalization of de Forest's 
``cc'' prescriptions~\cite{deForest} to be used with the 
weak neutral current. At low momentum transfer,
where relativistic corrections are --presumably-- not important, 
the FSI calculation should be used~\cite{MEC}. 
At TJNAF energies, though,
the nonrelativistic expansions in the FSI calculation are 
inadequate, while, on the other hand, rescattering effects are
expected to be suppressed  --- and therefore the PWIA calculation
should be preferred. In adopting this point of view, the $6\%$
effects at $\bq>0.7$ GeV/c, which, if taken at face value 
threaten the interpretability of PV deuterium measurements,
are not viewed as a fair measure of nuclear
uncertainty; rather, the FSI model should be considered 
not applicable for such kinematics~\cite{Amaro}.

The PWIA calculation, however, is
a semirelativistic, rather than a covariant calculation,
in that it treats the deuteron as a nonrelativistic bound state.
On the other hand, much work has been devoted to relativistic
treatments of $e-d$ scattering, in the sense of 
consistent nonrelativistic expansions including wave function
and boost effects~\cite{Ar1}, as well as 
covariant solutions of the bound state using
the Bethe-Salpeter equation~\cite{KT,HT,Umn} 
or its  ``spectator'' version~\cite{BGross,Arnold}.
Given the availability of such relativistic treatments, 
and the degree of precision required for strangeness studies,
the basic objective of this work is to calculate
the PV $e-d$ asymmetry in a covariant model. In doing so,
our motivation is different from most studies of relativistic 
effects, where observables exhibiting large such effects 
are sought after, in the hope to probe details of nuclear dynamics. 
In our case we wish to be as independent as
possible from nuclear dynamics. We wish to quantify 
that the {\em residual} relativistic effects which
are associated with a relativistically described deuteron bound 
state, and are, therefore, {\it not} included in the semirelativistic
PWIA, are small and, for the PV asymmetry, less than $1\%$ 
for the kinematics of interest. Since the high-momentum
 bound nucleons are suppressed due to Fermi motion, 
it is reasonable to expect such 
residual relativistic effects to be small.
Establishing such a result is important
for high $q^2$ quasifree experiments at TJNAF
with heavier nuclei as targets, for which a semirelativistic 
PWIA is possible but a covariant calculation is not 
(notwithstanding the fact that the effects reported here 
for the deuteron will underestimate those for such heavier 
nuclei due to the uncharacteristically loose binding 
($k_F\approx 60$ MeV/c) of the deuteron).

Although ultimately we would like to have a covariant formulation 
of PV $e-d$ scattering following the more fundamental approach of Hummel and 
Tjon~\cite{HT}, a simpler approach is taken here, following
that of Keister and Tjon~\cite{KT} and Beck et al.~\cite{BWA}. 
It amounts to a covariant, or ``relativistic'' (R) PWIA.
Unlike the semirelativistic, or factorized (F) PWIA,
in which the deuteron is treated nonrelativistically,
in this case the hadronic tensor does not factorize in the 
strong sense, i.e., cannot be written as a convolution of a 
single-nucleon tensor with a momentum distribution.  
This is due to the off-shellness of the 
propagating nucleon. By separating the forward (positive energy)
and backward (negative energy) components of the struck 
nucleon's propagator, we find a ``weak''
form of factorization, where the momentum distribution 
and the single-nucleon tensor become density matrices. 
As a byproduct of this construction, we can identify the 
{\it ad hoc} prescription in the factorized (F) 
PWIA of de Forest, simply, as a projection of the covariant hadronic 
tensor of the RPWIA onto the positive energy sector, which
shows the relation between off-shell and relativistic effects.
Since it is off-shell effects that 
spoil factorization  in RPWIA, we expect, and indeed 
observe numerically,  that the effects of 
the projection are minimal on the quasielastic peak, 
where kinematics are ``maximally on-shell''. 

The structure of this article is as follows. In Sect.~II
we discuss the relativistic calculation (RPWIA) used in this work. 
In Sect.~III we review the factorized PWIA,
and discuss the projection by which
the FPWIA is obtained from RPWIA. In Sect.~IV 
we show numerical results for the cross section and the 
PV asymmetry for the kinematics of interest, and discuss 
the role of the residual relativistic effects in the
interpretation of experiments designed to study
nucleon strangeness. Our conclusions
appear in Sect.~V and some details of the calculation
in the Appendices.

\section{the Hadronic Tensor in Relativistic PWIA}

\begin{figure}[htb]
\begin{center}
 \mbox{\epsfig{file=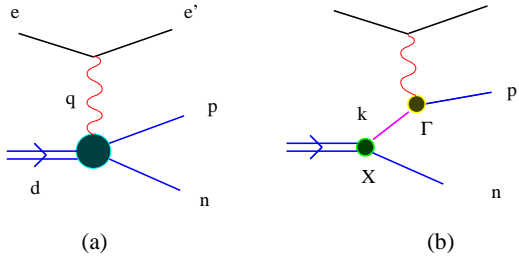,width=0.40\textwidth,angle=-90}}
\caption[dummy]{Deuteron electrodisintegration in PWIA.}
\label{kinem}
\end{center}
\end{figure}

The kinematics for deuteron electrodisintegration,
in PWIA and in the target rest frame, $d^\mu$ = $(M_d,\v 0)$, 
are shown in Fig.\ \ref{kinem}(a). The momentum transfer
is $q^\mu=(\omega,\v q)$, and, for the moment, we assume 
that the nucleon $p^\mu=(E_p,\v p)$ is detected in coincidence
with the electron $e'^{\m}=(\epsilon', {\bf u}')$, 
while the other nucleon $n^\mu$ = $(E_n,-\v k)$ is not. 
Here $d^2$ = $M_d^2$, $p^2$ = $n^2$ = $m^2$. The cross section
is obtained from
\be\label{s0}
d\sigma^{(h)} =  {\alpha^2\over q^4} 
                   { d{\v \e'} d{\v p} \ d{\v n} \over    (2\pi)^3 \md }
        {m_e^2\over \ep\ep'}{m^2\over E_p E_n}
 \delta^4( P_f -P_i)
l_{\mu\nu} H^{\mu\nu} \ ,
\ee
where $P_i^\m=d^\m+q^\m$, $P_f^\m=p^\m+n^\m$,
 $l_{\mu\nu}$ is the leptonic,
\be\label{lep_EM}
l_{\m\n} = \sum_{h'}
      <e',h'|\g_\m|e,h>^* <e',h'|\g_\n|e,h> \ ,
\ee 
and $H^{\mu\nu}$ the hadronic tensor
\be\label{pwia}
H^{\mu\nu} =  \b1sum_{\rm spins}
    <f|\hat J^\mu_{p}|i>^*<f|\hat J^\nu_{p}|i> \ ,
\ee
and where spin d.o.f. have been averaged over in the unpolarized
target ($|i>=|d>$) state and summed over in the final electron
and $np$ states.
The plane-wave
impulse approximation (PWIA) amounts to (I) 
using one-body current operators $J^\m$, (II) disregarding final-state
interactions, and (III) assuming that the
detected nucleon is identified with the 
(off-shell, $k^2$ $<$ $m^2$) nucleon  $k^\mu$ = 
$(E_k,\v k)$ struck by the virtual boson~\cite{deForest,pwba}. 
Under these assumptions the PWIA amplitude reduces 
to that of Fig.\ \ref{kinem}(b).  
The inclusive cross section in PWIA
with polarized electron beam and unpolarized target
is obtained by integrating over $d^3\v n$ and $d\Omega_k$ 
using the energy-momentum
$\delta$--function~\cite{yscaling}
\bea\label{s1}
	{ d^2\sigma^{(h)}\over d\O_e dw}
&=&
	{1\over (2\pi)^2}\; {1\over M_d} \ \sigma_{\rm M}  
         \int_{|y|}^Y d \k \; \k \; 
	\; {m^2\over q \sqrt{m^2+\k^2}} \nnn\\
&\times&  
	\sum_{i=p,n}\left( 
v_L W^i_L + v_T W^i_T + h\ v_{T'} W^i_{T'}\right) \\
&\equiv& \sigma_{\rm M} 
	\left[
v_L R^L+ v_T R^T + h\ v_{T'} R^{T'}\right]
        \nnn .
\eea
Here $\sigma_{\rm M}$ is the Mott cross section.
The limits of integration, $|y|$ and $Y$, 
as well as the responses $R$
are functions of $\bq$ and $\w$.~\footnote{
the $T'$ response is included anticipating the discussion of
parity-violation below; it is identically zero for purely 
electromagnetic (EM),
inclusive, unpolarized target scattering.}
 The lepton kinematics are 
included in the factors~\cite{Don_Rask}
\bea\label{xxx}
v_L &=& \left[ q^2\over {\v q}^2\right]^2,\; 
v_T={1\over 2}\left| q^2\over {\v q}^2\right|
+\tan^2{\theta\over 2}\nnn\\
 v_{T'} &=&\tan{\theta\over 2}
\sqrt{\le|q^2\over{\v q}^2\ri|+\tan^2{\theta\over 2}} \ .
\eea
The incoherent sum over nucleons in \equ(s1)
is a consequence of assumption (III) above~\cite{pwba}.
The respones $W^i$ are obtained as 
components of the hadronic tensor in \equ(pwia).

In relativistic (R) PWIA the hadronic tensor is
obtained from the amplitude in Fig.\ \ref{kinem}(b) 
\be\label{proton}
<f|\hat J^\mu|i> \sim \bar u(p) \; \Gamma^\mu \;
	{\s k + m\over k^2-m^2} \;
		\X_\n \; \x^\n \; v(n) \ .
\ee
Here $\Gamma^\mu$ is the half-off-shell nucleon vertex,
$\x_\n$ is the deuteron polarization vector and
$\X^\n$ is the d-p-$\bar{\rm n}$ vertex. Having applied charge 
conjugation to the on-shell nucleon $|n>$, rather than the off-shell
one $|k>$, $\X^\mu$ is obtained from the Blankenbecler 
and Cook~\cite{BC} vertex, $\V^\mu$, according to
\bea\label{tria}
\X_\n &=& \C^T \V_\n^T \C^T \nnn\\
	&=& \left. A\g_\n + \B \q_\n+
		    {m-\s k\over 2 m}	
	  \left[F\g_\n + \G \q_\n\right]
				\right. \ ,
\eea
with $C$ the charge conjugation operator, and
$\q^\nu=(n^\nu-k^\nu)/2$. The vertex functions $A,B,F,G$
depend on $k^2$ and can be written in terms of the
$^3S_1$ ($u$), $^3D_1$ ($w$), $^3P_1$ ($v_t$) and $^1P_1$ 
($v_s$) relativistic deuteron wavefunctions~\cite{KT,BGross}. 
In the on-shell matrix element $\bar u(k)\X_\nu v(n)$ 
the off-shell $F,G$ vertex functions would not contribute.
Squaring the amplitude in \equ(proton), 
summing over spins in the final state, 
and averaging over deuteron polarizations,
one obtains for the hadronic tensor in RPWIA
\bea\label{proton2} 
 H^{\mu\nu}_{\rm RPWIA} &=& 
{ -g_\rl + {d_\r d_\l/\md^2} \over
  3 (k^2-m^2)^2} \  {\rm Tr}\le\{
	{\s n- m\over 2m} \ \bar \X^\r \ (\s k + m)
\ri.\nnn\\ && \le. \qquad\qquad \times\;
\bar \Gamma^\m \ {\s p + m\over 2m}\ 
 \Gamma^\n \ (\s k + m)\  \X^\l \ri\} 
 \ ,
\eea
where $\bar \X_\n$ = $\gamma_0 \X^\dagger_\n \gamma_0$
and similarly for $\bar \Gamma_\m$.
Following Keister and Tjon~\cite{KT} 
we define auxiliary vectors $\U^\mu$, $S^\mu$ 
and scalars $\beta$, $T$ through
\bea
 \le(\s \U + \beta\ri) &=&
\left(-g_\rl + {d_\r d_\l/\md^2}\right)  \X^\l (\s n- m)  \bar \X^\r 
 \label{defb}\\
 \left( \s S + T \right) &=&
 {\s k + m\over 2m}  \le(\s \U + \beta\ri)
{\s k + m\over 2m}  \label{defs} \ , 
\eea
implying
\bea\label{ST}
 (2m)^2 S^\m &=& 2\left[ m\beta + 
  (k\c \U)\right] k^\m + (m^2-k^2) \U^\mu \nnn \\
	(2m)^2 T &=& (k^2+m^2)\beta + 2 m (k\c \U) \ .
\eea
We thus obtain
\be\label{proton3}
H^{\mu\nu}_{\rm RPWIA}
 =  {4m^2\over 3(k^2\!-\!m^2)^2}\  {\rm Tr}\left\{
	{\s S + T\over 2m}\ \bar \Gamma^\m\ {\s p + m\over 2m}\ 
	\Gamma^\n \right\} \ .
\ee
Expressions for $\U$, $\beta$, and the hadronic tensor
can be found in Appendix A. Equations (\ref{proton3})
and (\ref{ST}) provide a reference point for a comparison
with the factorized (F), or ``semirelativistic'',
PWIA, that is widely used in intermediate 
energy physics.

\section{reduction of RPWIA to FPWIA}

\subsection{Factorized PWIA revisited}

In FPWIA the cross section is written as
a convolution of a nonrelativistic momentum distribution
$\rho=\rho(\k)$ with a single nucleon cross 
section~\cite{deForest,yscaling}

\bea\label{conv}
	{d^2\sigma^{(h)}\over d\O_e d w} &=& 
2\pi\ \sigma_{M} \int_{|y|}^{Y}
                    d\k \ \k 
              \   \rho(\k) \  {m^2\over q\sqrt {m^2+\k^2}} \  \nnn\\
 &\times& \sum_{i=p,n}
      \left[ v_L \W^i_L + v_T \W^i_T + h\ v_{T'} \W^i_{T'} \right] \ ,
\eea	
The {\it single nucleon} responses 
$\W_{L,T,T'}$ are obtained as components of a 
covariant single nucleon tensor $\H^{\m\n}$,
describing the interaction of a moving, off-shell 
nucleon with the virtual photon. 
The most general form of the off-shell vertex 
has been obtained by Bincer~\cite{Bincer}. 
The special case of the half-off-shell vertex
$\bar u(p) \Gamma^\mu$ involves four independent vertex 
functions of two scalar variables, e.g.,  
$q^2$ and $k^2$, and collapses to the familiar form
involving just two form factors (Dirac, $F_1$, and Pauli, 
$F_2$) when operating on an on-shell spinor $u(k)$. 
The half-off-shell vertex functions of composite particles 
cannot be uniquely defined~\cite{Stefan,DP}. 
They can of course be computed in specific 
models~\cite{NK,others}, but then the whole reaction 
(i.e., the self energy and the d-p-$\bar {\rm n}$ vertex) 
should be, as well. Since such simple models are not realistic 
(i.e., they do not reproduce the observable on-shell 
form factors $F_1$ and $F_2$) it has become customary to resort to 
{\it ad hoc} prescriptions for treating the half-off-shell 
single nucleon tensor, the most widely used of which are
the ``cc'' prescriptions of de Forest~\cite{deForest}.
These prescriptions involve a set of operations regarding
\begin{enumerate} 
\item[I.] the wavefunction of the struck off-shell nucleon.
\item[II.] the form of the half-off-shell vertex.
\item[III.] the treatment of current conservation.
\end{enumerate} 
With respect to (I), the struck, off-shell
nucleon wavefuction is taken to be that of the 
corresponding ``would-be-on-shell''
spinor $u(\bk)$, with same spatial momentum $\v k$, but 
(positive) energy $\E=E_n=\sqrt{m^2+{\v k}^2} > E_k$. As we discuss 
in the next section, by reformulating (I) in a more 
transperent fashion, it is this approximation that 
differentiates between RPWIA and FPWIA.

With respect to (II) the following two vertices are employed
\bea\label{ccvertex}
\Gamma^\m_2 &=& F_1\gamma^\m + i{q_\n\over 2m}\sigma^{\m\n}F_2\nnn\\
            &\hookrightarrow& (F_1+F_2)\gamma^\m -{(k+p)^\m \over 2m}F_2
              + \gamma^\m F_2{(\s k -m)\over 2m}\nnn\\
\Gamma^\m_1 &=& (F_1+F_2)\gamma^\m -{(\bar k+p)^\m\over 2m}F_2\ ,
\eea
where the second equation in (\ref{ccvertex}) is to be understood 
in the sense of the Gordon decomposition of $\bar u(p) 
\Gamma^\m_2$. In the $\Gamma_1^\m$ vertex appears the 
auxilliary four-vector $\bk^\m \equiv(\E,\vk)$, 
instead of $k^\m=(E_k,\vk)$. In the on-shell case, the Gordon
decomposition of  $\bar u(p) \Gamma^\m_2 u(k)$ is complete 
and the $\Gamma^\m_1$, $\Gamma^\m_2$ vertices
matrix-element-equivalent. 
The difference between results obtained using
the two forms of the vertex is interpreted as 
an estimate of the theoretical uncertainty induced 
in the modeling of the reaction because of our
inability to calculate reliably the half-off-shell 
vertex functions from first principles~\cite{deForest,Juan}. 
Thus, under I and II, the single nucleon tensor reads
\be\label{snt}
	\H^{\m\n}_{\rm FPWIA} = \tr\left( \z(\bk) \;\bar \Gamma^\mu_{1,2}\; 
	\z(p)\; \Gamma^\n_{1,2}\right) \ ,
\ee
where (I) has resulted in the positive energy projector with 
the auxilliary four-vector $\bk$.

With respect to III, one notices that, 
in the half-off-shell case, the one-body
current is not conserved~\cite{NK}, 
since $q_\m \bar u(p) \gamma^\m$ = $\bar u(p) (m-{\s k})$.
The de Forest ``cc'' precriptions enforce current conservation 
${\v J}\cdot{\v q} = J_0 \w$ by eliminating the third component 
of the current in favor of the charge density, where $\v q$
defines the z-axis. Alternatively, one may choose to
eliminate the charge in favor of $J_3$, or not enforce current
conservation at all. In Ref.~\cite{Juan} it was shown that the 
treatment of current conservation induces, especially in
high missing momentum kinematics in $(e,e'p)$ processes,
a substantial uncertainty. Recently, the three ways of treating
current conservation have been shown to be equivalent to specific 
gauge choices within a family of covariant gauges~\cite{gauges}.
Thus, the RPWIA approach described here is covariant, 
but not gauge invariant.

 \subsection{Strong {\it vs.} Weak Factorization}

The question we address here is under which set of 
operations (related to assumption (I) above) 
is the FPWIA of Eqs.~(\ref{conv}) and (\ref{snt})
obtained from the RPWIA of Eqs.~(\ref{s1}) and (\ref{proton3})
\be\label{redu}
	H^{\m\n}_{\rm RPWIA}
	\stackrel{?}{\rightarrow}
	(2\pi)^3\md\;\rho\;\H^{\m\n}_{\rm FPWIA} \ .
\ee
We start by observing that from \equ(ST) one has
\be\label{fact1}
	\thru S + T = 2\left[ m\beta + \prod(k,\U)\right]
	{\thru k+m\over 4m^2} + 
        \left.m^2-k^2\over 4m^2\right.\left[\thru \U -\beta\right] \ .
\ee
Since $\U^\n$ is obtained from the deuteron vertex functions 
[cf. \equ(defb)], the second term in \equ(fact1) does not allow to 
factor all dependence on details of the deuteron 
break-up amplitude out of the
trace of \equ(proton3), thus defining a
``single nucleon'' tensor. However, if one ignores the
off-shellness of the struck nucleon and 
sets $k^\m\rightarrow \bk^\m$ in \equ(ST), one obtains
\bea\label{fact2} 
	4m^2(\thru S + T) \rightarrow 2\Bigl[m\beta + \prod(\bk,\U)\Bigr]
				(\thru \bk+m) \ .
\eea
Comparison with Eqs.\ (\ref{s1}), (\ref{proton3}), 
(\ref{conv}) and (\ref{snt}) shows 
that the cross section has indeed the factorized form 
of the ``semirelativistic'' FPWIA, provided that the
nonrelativistic momentum distribution $\rho$  
is related to the quantity $\left[m\beta + 
\prod(\bk,\U)\right]/\left[12\pi^3 M_d(k^2-m^2)^2\right]$. 
We shall return to this later on. The relevant 
observation at this point is that the RPWIA is 
{\it not} factorizable, unless the off-shellness 
of the struck nucleon is ignored.~\footnote{
analogous observations can be made for the case of
deep inelastic scattering processes~\cite{Thomas}.}
To probe this issue further, we use the familiar 
technique of interpreting the propagation 
of the off-shell struck nucleon as a superposition
of positive and negative (on-shell) energy states
(see, e.g., Ref.\ \cite{Compton}, or Arnold et 
al.~\cite{Arnold}, where a similar 
formalism is developed for elastic $e-d$ scattering)
\be\label{split}
	\z(k) = \fp \left. {\s k}_+ +m\over 2m\right. + 
                \fm \left. {\s k}_- -m\over 2m\right. \ ,
\ee
where 
\be\label{explain}
	f_\pm   = (E_k \pm \E)/ 2\E \ , \; \;{\rm and}\;\;\;
        k_\pm^\m  =  (\E, \pm\vk) \ .
\ee
The decomposition, \equ(split), is frame-dependent. 
Notice, in particular, that in the deuteron rest frame
the on-shell states are the struck and spectator nucleon,
respectively, i.e., $\kp^\m=\bk^\m$ and 
$\km^\m = n^\m$. We now write
\equ(split) in terms of the direct products of these
on-shell states
\be\label{direct}
	\z(k) =   \fp \sum_s \us \bus + \fm \sum_s \vs \bvs \ ,
\ee
and perform this decomposition on \equ(proton2). In the 
shorthand notation $\chi^s_+\equiv\us$ and $\chi^s_-\equiv\vs$,
the trace of \equ(proton3) now reads (Dirac indices explicitly
shown and implicitly summed over)
\bea\label{factor1}
   & \sim& \sum_{\rm s,r}\sum_{a,b=\pm} f_a f_b 
        \left[\thru\U + \beta\right]_{ij} 
        \left[\chi^s_a\right]_j 
	\left[ \bar\chi^s_a\right]_k 
        \left[\bar\Gamma^\m\right]_{kl}
	\left[\thru p + m\right]_{lm}
\nnn\\&&\qquad\qquad\qquad\qquad\times
       \left[\Gamma^\n\right]_{mn}
        \left[\chi^r_b\right]_n
	\left[\bar \chi^r_b\right]_i \ .
\eea
By inspection, this can be cast in the form
\be\label{density} 
	H^{\mu\nu}_{\rm RPWIA} = (2\pi)^3 \md 
                      \sum_{\rm spins} \ \sum_{a,b=\pm} 
                        \rho_{ab;rs}\ \H^{\m\n}_{ab;rs} \ . 
\ee
A {\em weak}  form of factorization, therefore, 
still holds, with momentum distribution 
$\rho_{ab;rs}$ now being a density matrix in both 
spin and energy projection indices
\be\label{gen_mom}
	\rho_{ab;rs} = f_a f_b {m^2\over 6\pi^3\md(k^2-m^2)^2}
	            \tr\left\{ \chi^s_a \bar\chi^r_b 
		           \;{\thru\U + \beta\over 2m} \right\} \ ,
\ee
and similarly for the single-nucleon tensor
\be\label{gen_snt}
	\H^{\m\n}_{ab;rs} = \tr\left\{ \chi^r_b  \bar \chi^s_a \;
			 \bar\Gamma^\m \; \z(p) \; 
			  \Gamma^\n\right\}  \ .
\ee
Using $\vs = \g^5\g^0 \us$ and~\cite{thesis} 
\be\label{proj}
	\us \; \bur = \z(\bk) \; 
       {\delta_{s,r} + \g^5\thru\xi_{s,r} \over 2} \ ,
\ee
where $\xi^\mu_{s,r}$ = 
$\bar u({\v k}, r) \g^5\gamma^\mu  u({\v k}, s)$,
we find
\bea
	\rho_{\pm\pm;rs} &=& \delta_{s,r}\;  
           f_\pm^2 \left. {\prod(k_\pm,\U) \pm m\beta
       \over 12\pi^3\md(k^2-m^2)^2}
                            \right. 
\label{rho_traces}\\
	\rho_{+-;rs} &=& \fp\fm \left. 
\E \prod(\xi_{s,r},\U) - \left[
\prod(\kp,\U)+m\beta \right] \xi^0_{s,r} 
		  \over
12\pi^3\md(k^2-m^2)^2\right.
\label{rho_traces_off} \nnn\\
&=&\rho_{-+;rs} \ \ .
\eea
Thus, the diagonal terms $(+,+)$ and $(-,-)$ are diagonal
in spin indices as well. Setting
\be\label{shift}
   \rho_{\pm\pm;rs}  =    \rho_{\pm\pm} \  \delta_{r,s}, \;\;
\sum_{\rm spins} \delta_{r,s} \ \H^{\m\n}_{\pm\pm;rs} 
 =  \H^{\m\n}_{\pm\pm} \ ,
\ee
we obtain for the diagonal contributons
\be\label{reduction}\sum_{\rm spins}
   \rho_{\pm\pm;rs}\ \H^{\m\n}_{\pm\pm;rs} 
   = \rho_{\pm\pm} \ \H^{\m\n}_{\pm\pm} \ ,
\ee
where we make use of $\sum_s  \xi^\mu_{s,s} = 0$
to identify  $\H^{\m\n}_{++}$ = $\H^{\m\n}_{\rm FPWIA}$ 
[cf.~\equ(snt)], $\rho_{++}$ =  
$\fp^2[m\beta +\prod(\bk,\U)]$ $/12\pi^3 M_d( k^2-m^2)^2$,
and analogously for the $(-,-)$ contribution. The 
observation made in \equ(fact2) can thus be 
formally stated as follows: the first step
(actually the {\em only} step, as shown below) 
required for the reduction to FPWIA [cf. \equ(redu)] 
is a projection to $(+,+)$ energy sector.
With respect to the off-diagonal (in energy indices, $a\ne b$)
contribution, notice that the relevant 
momentum distribution does not appear to be
diagonal in spin space either. The second term 
in \equ(fact1), which, as noted above, spoils strong 
factorization, is an off-diagonal energy $(+,-)$ term, 
as suggested by writing
\be\label{poff}
m^2-k^2 = \E^2-\Ek^2 = -2\fp\fm\left[m^2+\prod(\bar k,\eta)\right] .
\ee
This is verified by explicit evaluation in Appendix B.
 
\subsection{Nonrelativistic Momentum Distribution}

We now address the relationship between $\rho_{++}$
and the familiar, nonrelativistic deuteron 
momentum distribution $\rho(\k)$ $\sim$ 
$\left[u^2(\k)+w^2(\k)\right]$. In terms of the 
wavefunctions $u$, $w$, $v_s$ and $v_t$, the
vertex functions in \equ(tria) are given by~\cite{BGross}
\bea\label{rend}
A &=& {\cal N}
       \left[ u - {1\over \sqrt{2}}w +{m\over \k}\sqrt{3\over 2}v_t
       \right]  \nnn\\
B &=& {\cal N}
     \left[ {m\over \E+m}u + 
      {m(2\E+m)\over \sqrt{2}\k^2}w
      +{m\over \k}\sqrt{3\over 2}v_t
      \right] \nnn\\ 
G &=& {2 {\cal N} m^2\over \md}
     \left[ {u\over\E+m} -
      {\E+2m\over \sqrt{2}\k^2} w   \right] 
     +2\pi\sqrt{6\md} {m^2\over \k}v_s \nnn\\
F &=&-2\pi\sqrt{2\md}\E\sqrt{3\over 2}{m\over\k}v_t \ ,
\eea
where ${\cal N}$ = $\pi\sqrt{2\md}(\E-\Ek)$.
Intuitively, we expect that $\rho_{++}$ reduces
to $\rho$ in the nonrelativistic limit, that is, by
ignoring the $P$ states, and, additionally,  making a 
$({\k}/m)^2$ expansion~\cite{spin96}. It 
turns out, however, that neither approximation 
is necessary, and, in fact, $\rho_{++}$ and 
$\rho$ are {\it identical}. To see this, first
notice that the overall contribution of the $P$ 
states must cancel out once the projection to the 
$(+,+)$ sector is made, since such states 
represent ``lower'' Dirac components $\psi^-$ of the 
deuteron wavefunction: from Eq. (38) in 
Ref.~\cite{BGross} we have
(in our notation) 
\bea\label{psis}
^3S_1,{}^3D_1 \; \in \; \psi^+ &\sim & \bar\chi_+ \prod(\X,\xi)v(n) \nnn\\
^3P_1,{}^1P_1 \; \in \; \psi^- &\sim & \bar\chi_- \prod(\X,\xi)v(n) \  .
\eea
By construction, the $(+,+)$ sector does not contain $\chi_-$ terms.
Thus, $\rho_{++}$ is oblivious to the $P$ states (we have
verified this numerically, as well), which can therefore be
neglected for the purposes of writing $\rho_{++}$ in terms of
wavefunctions. Notice that, although in this case $F\rightarrow 0$, 
there is still an apparently nonvanishing contribution
to $\rho_{++}$ from the ``off-shell'' vertex function $G$; 
however, this can be shown to vanish identically. 
For example, from Eqs.\ (\ref{scalar}) and (\ref{vector}) in the 
appendices,  and using \equ(rend),
we have for the $G^2$ terms in the deuteron rest frame 
\bea\label{gterms}
{\cal O}(G^2)
&\sim& m^2 + k^2
      +(k^2\!-\!m^2){\prod(\bar k,n)\over m^2}
         -2\prod(k,\bk)
\nnn\\ && \qquad  
       + 2\prod(k,n)-2\prod(k,n){\prod(k,\bar k)\over m^2}
\nnn\\
&=&  m^2+\Ek^2 + 2\Ek\E +(\Ek^2\!-\!\E^2){\E^2+{\k}^2\over m^2}\nnn\\
& &  \qquad - {2\over m^2}(\E^2\Ek^2-{\k}^4)-2\E\Ek  +3{\k}^2 \nnn\\
&=&  m^2-2{\E^4\over m^2} + \E^2 + 3{\k}^2 +2{{\k}^4\over m^2}\nnn\\
&=&  0 \ \ ,
\eea
and similarly for the $AG$ and $BG$ terms. Thus, 
once the $P$ states have been shown to decouple, 
the nonvanishing contribution comes from the $A$ 
and $B$ vertex functions alone. From \equ(scalar) and \equ(vector)
with $\D^2=-\k^2$, we find
\bea\label{epit}
   m\beta + \bk\c\U &=& A^2\left(3m^2 + 3\prod(n,\bar k)- 
2\prod(N,\bar k)   \ri)\nnn\\
    &+&B^2{\D^2\over m^2}\left(m^2-\prod(n,\bar k)\right)
-2AB \left(\prod(\D,n)-\prod(\D,\bar k)\ri) 
 \nnn\\
&=& A^2 (6m^2 + 4{\k}^2) + 4 A B {\k}^2 + 2B^2{ {\k}^4\over m^2}
    \ .
\eea
To leading nonrelativistic order we find from \equ(epit) 
and \equ(rend) (without $v_t$!)
\bea\label{leading}
m\beta + \bk\c\U &=& {\cal N}^2\Bigl[ 
         6m^2 \le(u-w/\sqrt 2\ri)^2 
     + 9 m^2 w^2 \nnn\\
&&\qquad\qquad  + 12m^2\le( u-w/ \sqrt{2}\ri) 
{ w/ \sqrt{2}} \Bigr] \nnn\\
&=& 6{\cal N}^2 m^2 \left(u^2 + w^2\right)  \ .
\eea
After some suppressed algebra we find that retaining all 
terms modifies \equ(leading) multiplicatively, by ${\E^2/m^2}$.
Thus, we finally obtain
\bea\label{allord}
             m\beta + \bk\c\U 
        &=&  6{\cal N}^2 \E^2 \left(u^2 + w^2\right) 
\nnn\\
               \Rightarrow \rho_{++}
        &=& \fp^2\left. m\beta + \prod(\bk,\U)\over
                      12\pi^3 M_d( k^2-m^2)^2\right.
\nnn\\
        &=&    { 6 {\cal N}(\E+\Ek)^2\E^2(u^2+w^2)
                          \over 
                48 \pi^3 \md \E^2 (k^2-m^2)^2} \nnn\\
        &=&     { u^2+w^2 \over 4\pi} \ ,
\eea
where we have used \equ(explain) and $k^2-m^2$ = $\Ek^2-\En^2$.
 $\rho_{++}$ is precisely the nonrelativistic momentum 
distribution, apart from an overall normalization, since, 
when $P$ states are present, $\rho_{++}$ is not
suitably normalized, i.e., $\int_0^\infty d\ka \ka^2 [u^2+w^2] < 1$.  
 Before proceeding to the next section, 
two comments seem appropriate. First, it is 
important that $\rho_{++}$ contains 
$\bk\c\U$, rather than $k\c\U$. This can be seen 
by observing that the leading order coefficient 
of the  $B^2$ term must be  
$\sim {\k}{}^4$, otherwise $\rho_{++}$ diverges 
as ${\k}\rightarrow 0$ [cf. \equ(rend)]. Using
$k\c\U$ would add to the RHS of \equ(epit) 
a term $B^2{\k}{}^2\E(\Ek-\E)/m^2$, which,
because of the deuteron binding energy $b$, would diverge
as ${\k}\rightarrow 0$, since $\E-\Ek = |b| +{\cal O}({\k}{}^2/m^2)$.
Second, we emphasize that Eqs.\ (\ref{gterms}) and (\ref{epit})
are only valid {\em after} the overall contribution of $P$ states has 
cancelled out. In other words, the $F$ and $G$ vertex functions do
contribute in $\rho_{++}$ when the $P$ states are present; their
contribution, though, serves to cancel that of the $P$ 
states from the $A$ and $B$ terms.


\section{Validity of the Reduction}

\subsection{Kinematical considerations}

In the previous section it was shown that the
de Forest type of factorized PWIA is obtained from 
the RPWIA by projecting onto the the 
positive energy sector of the propagating
 struck nucleon. The legitimacy of this 
projection depends on $\fp\simeq 1$, $\fm << 1$: 
the more off-shell is the struck nucleon, the less
valid the projection.
Energy/momentum conservation at the d-p-$\bar {\rm n}$
vertex in the deuteron rest frame implies that the 
struck nucleon is off-shell by
\bea\label{generic}
\E - \Ek &=& 2\sqrt{m^2+{\v k}^2} - \md \nnn\\
         &=& |b| +{{\v k}^2\over m} +   \dots \ \ ,
\eea
where the ellipses denote higher order relativistic terms.
The binding effect is negligible for the deuteron ($|b|\approx 
2.22$ MeV) and $|{\v k}|$ is effectively restricted by the small
deuteron Fermi momentum ($k_F\approx 60$ MeV/c), resulting in
a rapid falloff of the momentum distribution $\rho(|{\v k}|)$.
The projection seems, therefore, meaningful (at least in the 
deuteron rest frame). Some caution must be exercised, however,
for two reasons. First, there may be cases, where, despite
$\fp\simeq 1$, backward propagation is essential. 
An example is low energy Compton scattering from, e.g.,
a proton, where, to leading order, forward propagation cancels 
between the $s$ and $u$ channels and the classical (Thomson) 
limit is entirely due to  the backward propagation 
terms~\cite{Compton}. Second, the applicability of 
the Fermi motion argument depends
on the kinematics (choice of energy transfer $\w$ for fixed momentum 
transfer $\bq$), which determines the integration region for 
inclusive scattering [cf. \equ(conv)]. Overall energy/momentum 
conservation implies~\cite{yscaling,thesis} for the integration limits
$k_{\rm min} = |y|$, $k_{\rm max}=Y$ 
\be\label{ylimit}
y(\bq,w)= (\md+\w)\sqrt{{1\over 4}-{m^2\over s}} -{\bq\over 2} \ ,
\ee
and $Y=y+\bq$, with $s\equiv (\md+\w)^2 - \bq^2$. On the quasielastic
peak ($\w=\w_{qe}$) 
\bea\label{peak}
        \md + \w_{qe} &=& \sqrt{m^2+\bq^2} + m \nnn\\
\Rightarrow \w_{qe}   &=& {q^2\over 2m}  + {\cal O}\le(|b|\ri)
\eea
the lower limit $|y|=0$, and, therefore, the contribution
of small momentum components $\k$ saturates
the nuclear response in \equ(conv). Higher momentum 
components contribute as well, however, their importance (as
weighted by the rapidly falling momentum distribution), is 
negligible. Away from the quasielastic peak, though, $y\ne 0$.
As an extreme example consider the situation at threshold, where
$s=4m^2$, and thus
\bea\label{thr}
        \md + \w_{thr} &=& \sqrt{4m^2+\bq^2}  \nnn\\
\Rightarrow \w_{thr}   &=& {q^2\over 4m}  + {\cal O}\le(|b|\ri) \ .
\eea
In this case $|y|=\bq/2$, and the struck nucleon
will be off-shell by $1-\Ek/\E \ge (2-\md/\sqrt{m^2+y^2})$,
which, near threshold, translates into 24\%, 63\% and
115\% for typical TJNAF processes with $\bq$=1, $\bq$=2 and  
$\bq$=4 GeV/c, respectively. 
 
\subsection{Electromagnetic responses and the cross section}

\begin{figure}[htb]
\begin{center}
 \mbox{\epsfig{file=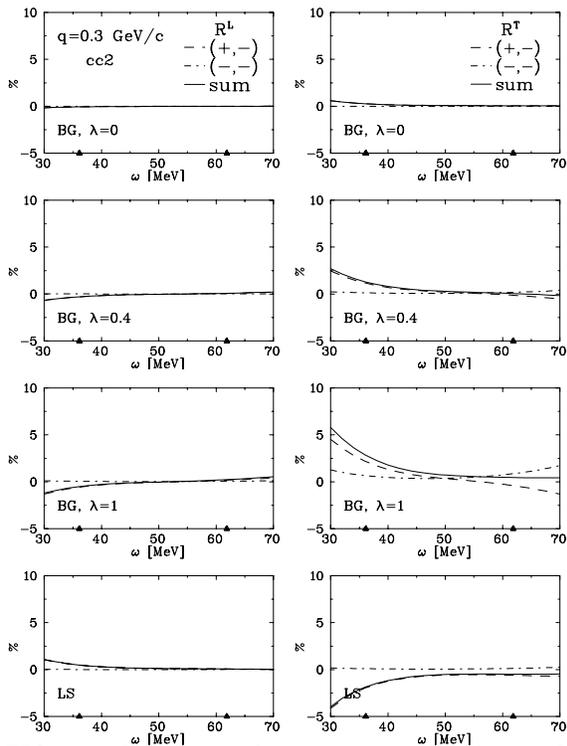,width=0.45\textwidth,angle=0}}
\caption[dummy]{ Percentage decomposition of inclusive electromagnetic 
                 responses (left panels, $R_L$, right panels, $R_T$) 
                  using the $\Gamma^\m_2$ vertex, 
                 at $\bq=300$ MeV/c. Dashes: $(+,-)$ contribution; 
                 dot-dashes: $(-,-)$ contribution; solid: total
                 contribution of negative energy components, i.e.,
                  $1-R^{++}/R$. Results shown for different choices 
                 of deuteron vertex functions. The solid triangles 
             mark the quasifree region  [cf. \equ(ridge)].
}         
\label{pr2_300}
\end{center}
\end{figure}

\begin{figure}[htb]
\begin{center}
 \mbox{\epsfig{file=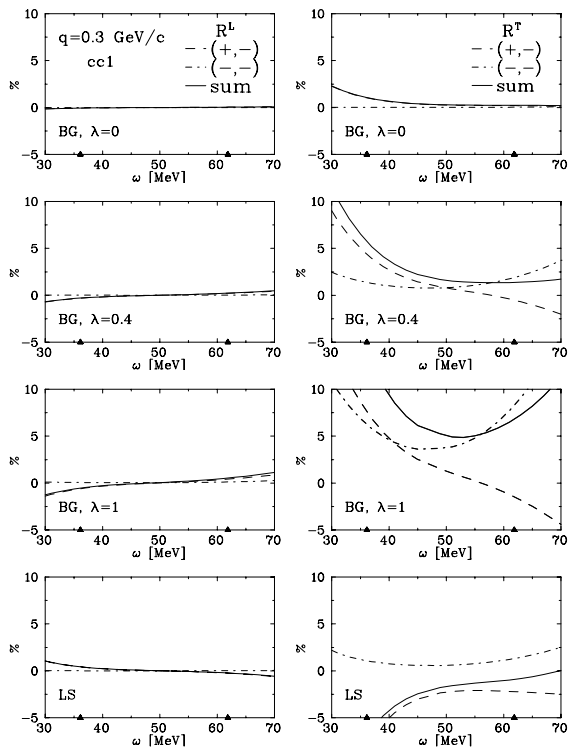,width=0.45\textwidth,angle=0}}
\caption[dummy]{As in Fig.\ \ref{pr2_300}, but with the $\Gamma^\m_1$
                vertex.}
\label{pr1_300}
\end{center}
\end{figure}

\begin{figure}[htb]
\begin{center}
 \mbox{\epsfig{file=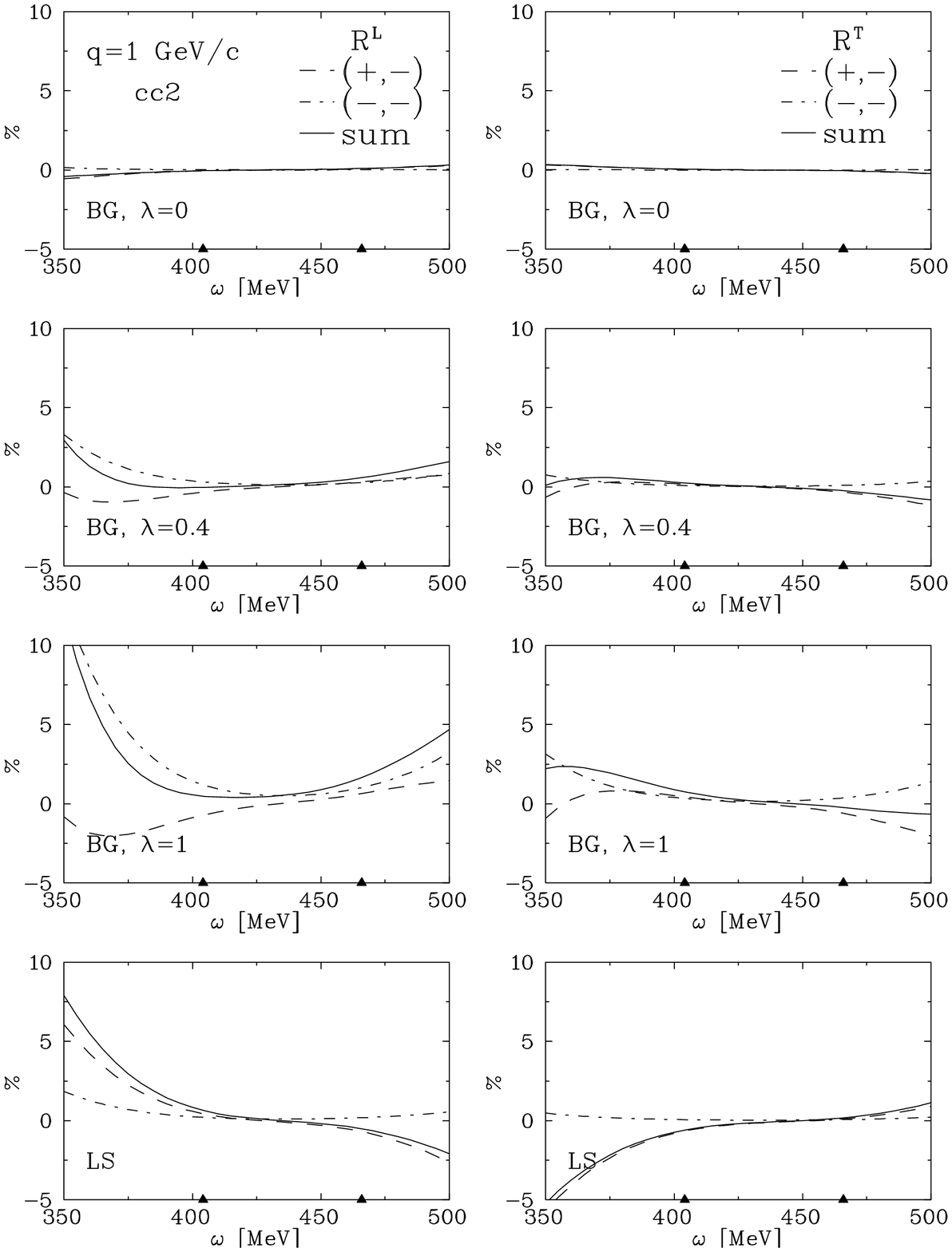,width=0.45\textwidth,angle=0}}
\caption[dummy]{As in Fig.\ \ref{pr2_300}, but 
                at $\bq=1$ GeV/c.
}
\label{pr2_1000}
\end{center}
\end{figure}

\begin{figure}[htb]
\begin{center}
 \mbox{\epsfig{file=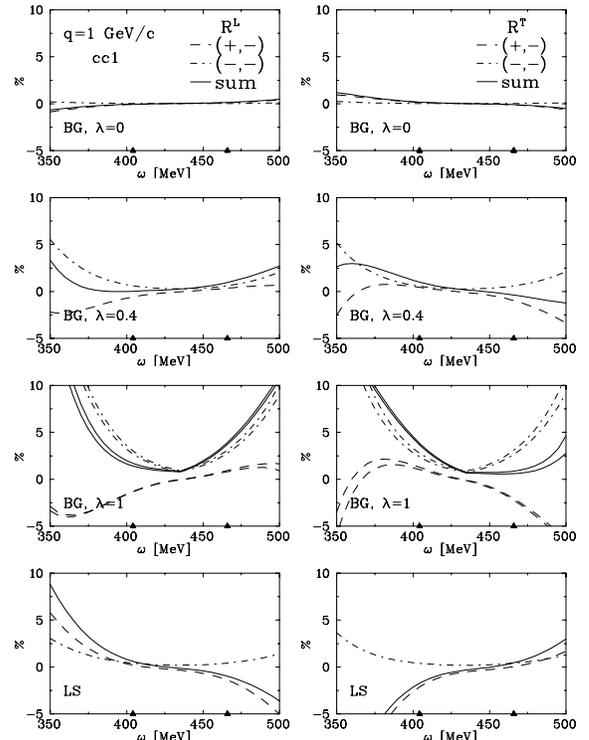,width=0.45\textwidth,angle=0}}
\caption[dummy]{As in Fig.\ \ref{pr2_1000}, but with the $\Gamma^\m_1$
                vertex. In the $\lambda=1$ panels results are shown 
               with and without the modification of the 
             on-shell form factors
               according to \equ(mod_naus).}
\label{pr1_1000}
\end{center}
\end{figure}

\begin{figure}[htb]
\begin{center}
 \mbox{\epsfig{file=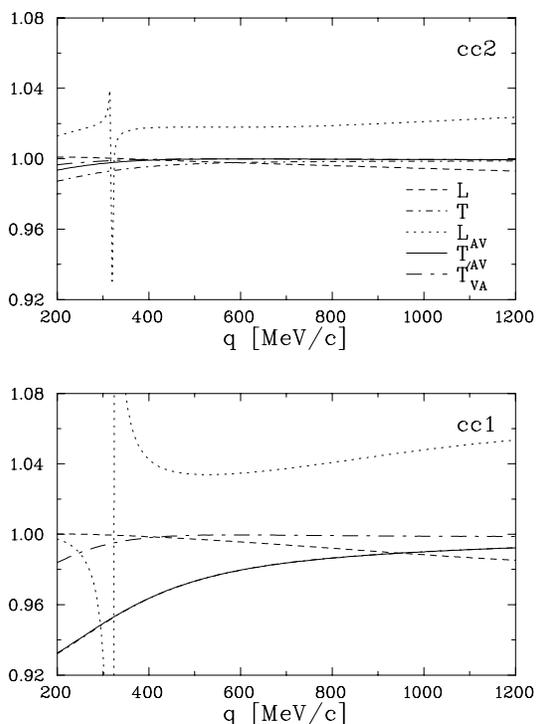,width=0.45\textwidth,angle=0}}
\caption[dummy]{The ratio $R^{++}/R$ of electromagnetic
                ($L,T$) and parity-violating 
                ($L_{AV},T_{AV},T'_{VA}$) responses
                accessible in inclusive $d(\vec e,e')$
                scattering, for quasifree kinematics ($\w=\w_{qe}$)
                as a function of momentum transfer $\bq$.
                The $\lambda=1$ Buck-Gross deuteron 
                vertex functions are used. Results are displayed
                for both cc1 and cc2
                prescriptions for the single-nucleon tensor 
                (the $T$ and $T_{AV}$ curves cannot be disentangled
                in the cc1 panel).
               }
\label{resp-wqe}
\end{center}
\end{figure}

The effects of the projection on the 
longitudinal ($R_L$) and transverse ($R_T$)
responses entering the inclusive cross section
can be seen in Figs.\ \ref{pr2_300}--\ref{resp-wqe}.
The responses are decomposed into $(+,+)$, $(+,-)$
and $(-,-)$ contributions, after carrying out the 
spin summation in \equ(density). By  $(+,-)$
we denote the total off-diagonal contribution, i.e.,
 $(+,-)$ plus  $(-,+)$ in the notation of \equ(density).
Unlike the $(+,+)$ and $(-,-)$ contributions which are 
always positive in the case of the electromagnetic responses,
the off-diagonal ones appear with either sign.
For numerical evaluation, it is easier to  project directly
from \equ(proton3) by applying the decomposition 
(\ref{split}) to \equ(defs), and then use
$\fp-\fm=1$ to arrange the resulting expression in
$\fp^2$, $\fp\fm$ and $\fm^2$ terms. We have verified that
this is identical to the direct trace evaluation 
of \equ(density) described in Sect.~III and Appendix B.

The residual relativistic effects we are interested in
correspond to the difference between RPWIA and FPWIA results.
From the discussion of the previous section follows that
the $R^{++}$ responses are identical with the FPWIA ones 
modulo an overall normalization, since $\int d\ka \ka^2 (u^2+w^2) < 1$
when $P$ states are present, while in FPWIA one uses wavefunctions
normalized to unity. This results to a 
maximum 1.4\% effect (with the $\lambda=1$ vertex functions, 
see below). In the case of the 
PV asymmetry, where such 1\% differences are important, 
this renormalization is irrelevant, as it cancels 
in the ratio of cross sections. Thus, in the following we 
use the terms $(+,+)$--projected and FPWIA interchangeably. 

We present results for $\bq=0.3$ GeV/c, representative
of a low energy experiment such as the SAMPLE
project, and $\bq=1$ GeV/c, representative 
of a medium energy experiment at TJNAF.
Current conservation has been enforced (``cc'') 
by eliminating $J^3$, as discussed in Sec.~III.
For the on-shell form factors appearing in the electromagnetic
vertices $\Gamma^\mu_{1,2}$ [cf. \equ(ccvertex)],
we assume that they will be experimentally well known
by the time the PV measurements will have been completed, 
and thus use the parametrization in Refs.~\cite{Galster,rev} 
rather than model calculations.
In this parametrization $G_E^p=(1+4.97\t)^{-2}$, 
$G_M^{p,n}=\mu_{p,n} G_E^p$ and 
 $G_E^n= -\t G_M^n (1+5.6\tau)^{-1}$,
with $\t=|q^2|/4 m^2$ and where $\mu_p=2.79$, $\mu_n=-1.91$ are
the proton and neutron anomalous magnetic moments, respectively.
For the four vertex functions in the d-p-$\bar {\rm n}$ vertex
we use the spectator equation solutions obtained by 
Buck and Gross~\cite{BGross}. These solutions are 
labelled by a parameter $\lambda$ that
interpolates between pseudovector ($\lambda=0$)
and pseudoscalar ($\lambda=1$) $\pi NN$ coupling
\be\label{inter}
 - i g_{\pi} \bar\psi \left[ \lambda - i (1-\lambda) {{\s \partial}\over 2m}
          \right ]\gamma_5 \phi \psi \ .
\ee
From the interpretation of the $P$ states as lower Dirac 
components of the deuteron wavefunction, one expects
that pseudocsalar (ps) coupling would enhance the role of 
the $P$ states, since $\gamma_5$, being off-diagonal, mixes 
lower and upper components. As discussed in Ref.~\cite{BGross} 
this is indeed the case, and the importance of $P$ 
states increases rapidly between $\lambda=0$ (pv) 
and $\lambda=1$ (ps). 

Commensurate, looking at the ``BG'' panels
in Figs.\ \ref{pr2_300}--\ref{pr1_1000}, we see that 
the difference between FPWIA and RPWIA, i.e., 
the error induced by projecting onto the $(+,+)$ sector, 
significantly increases with $\lambda$. 
It is well known, however, that the ps and pv 
$\pi NN$  Lagrangians are related via a field redefinition
(to lowest order in $\lambda$, at higher orders the field 
transformation generates contact terms, as well). According 
to basic field theory properties, the amplitude 
of the overall on-shell process should not depend on 
$\lambda$~\cite{unit,DP}. The fact that it 
actually does, partially reflects 
the inconsistency of computing the d-p-$\bar 
{\rm n}$ vertex functions in a specific model 
(both the on-shell, $A$, $B$ and off-shell, $F$, $G$) 
ones, while not using the same model for the vertex 
functions of the photon-nucleon vertex. 
In particular, it is inconsistent
to use the off-shell vertex functions $F$ and $G$ in the
d-p-$\bar {\rm n}$ vertex of \equ(tria), while ignoring 
altogether the two extra vertex functions entering the 
half-off-shell $\gamma NN$ vertex. As a side remark,
one notices that, while realistic 
(in the sense that deuteron properties can be reproduced 
using them) model solutions have been obtained for
 $F$ and $G$, first-principles 
calculations of the off-shell electromagnetic vertex 
functions are indicative rather, than realistic~\cite{NK}.
Conceivably, this reflects the fact that in the deuteron
off-shell vertex one needs the {\em constituent}  
off-shell, while, in the  $\gamma NN$ off-shell vertex, 
one needs the {\em bound state} itself 
off-shell, which is, apparently, more difficult to describe.

It appears, therefore, reasonable to ignore the off-shell 
vertex functions in the d-p-$\bar {\rm n}$ vertex, as long as
only on-shell electromagnetic form factors are used.
This point of view has been adopted, e.g., in Ref.~\cite{BWA},
where a parametrization of the deuteron vertex by Locher and
Svarc~\cite{LS} using only the $A$ and $B$ vertex functions was
employed. Since our primary motivation in this work is to 
investigate as widely as possible the extent to which 
effects associated with the relativistic aspects of the 
deuteron vertex modify the factorized PWIA, we will, 
nevertheless, use the off-shell deuteron vertex functions. 
In an attempt to partially alleviate the inconsistency of 
this scheme, we also show some results, whereupon we still
use only two vertex functions for the $\gamma NN$ vertex,
but modify the on-shell form factors according to
\be\label{mod_naus}
F(q^2,k^2) = F_{\rm Galster}(q^2) \times  
\le[ F_{\rm ps}(q^2,k^2)\over F_{\rm ps}(q^2,m^2)\ri ]  \ ,
\ee
allowing for a dependence on the other scalar variable, $k^2$.
For the modification factor we use the one-loop calculation
with pseudoscalar (ps) $\pi NN$ coupling  of Naus and Koch~\cite{NK}.
Admittedly {\it ad hoc}, this procedure (first adopted 
in Ref.\ \cite{Pollock}) incorporates, nevertheless,
what has been shown to be  -- within a toy model -- 
the prominent feature of the EM half-off-shell vertex. 
For consistency, we only apply this modification scheme with
the pseudoscalar ($\lambda=1$) deuteron vertex functions. 
For comparison purposes, we also show some results in 
the Locher-Svarc scheme (labelled ``LS''). We find, 
in general, that the effects of the projection are large 
in this scheme, comparable in magnitude to those with 
the $\lambda =1$ Buck-Gross vertex functions.

Looking at Figs.\ \ref{pr2_300} and \ref{pr1_300}, one sees that
at the low momentum transfer value ($\bq=300$ MeV/c),
the effects of neglecting negative ($-$) components 
in the longitudinal response, $R_L$, are negligible
for all energy transfer values $\omega$  
considered, independently of the deuteron vertices
and form of the EM current operator used.
For the transverse response, $R_T$, more significant
effects are seen, particularly with the $\Gamma^\mu_1$ 
vertex. Specifically, with the $\Gamma^\mu_2$ 
vertex (Fig. \ref{pr2_300})
the effects increase with $\lambda$ but are still below 2\% 
inside the quasielastic ridge
\be\label{ridge}
\vert w-w_{qe}\vert < \sqrt{2} k_F \bq/2\sqrt{m^2+\bq^2} \ .
\ee
For the  $\Gamma^\mu_1$ vertex, however, 
5-10\% effects are seen (Fig.\ \ref{pr1_300}), 
especially with the $\lambda=1$ vertex functions. 
In all cases, the effects are minimal 
for quasifree kinematics, as anticipated 
from the discussion in the previous section.
At this momentum transfer value, we find that 
a modification according to \equ(mod_naus) does not
produce an observable effect in the
the $\lambda=1$ panels of Fig.\ \ref{pr1_300}.

Unlike the case at the low momentum trasnfer, 
at $q = 1$ GeV/c (Figs.\ \ref{pr2_1000} and \ref{pr1_1000}), 
$R_L$ is affected by the projection more than $R_T$.
Inside the quasifree region, though, the effects 
do not exceed 3\% for either $R_L$ or $R_T$.
Thus, although the behavior of $R_T$ is qualitatively 
similar to what happens at $\bq=300$ MeV/c,
the effects -- particularly with cc1 -- 
are, within the respective quasifree regions,
{\em smaller} than at $\bq=300$ MeV/c.
This behavior does not contradict the remarks 
following \equ(thr), because threshold is close to 
the quasifree region at $\bq=300$ MeV/c (where 
$w_{thr}=24$ MeV), but not at  $\bq=1000$ MeV/c, where 
$w_{thr}=250$ MeV; that large effects are expected towards
threshold can be seen by extrapolating 
the $\lambda=1$ cc1 $R_T$ panel in Fig.\ \ref{pr1_1000}.
This behavior in the quasifree region
verifies that the {\em residual} relativistic effects
we describe are rather independent of the 
momentum transfer $q^2$. If anything, they appear
more important for low $q^2$ values, as can be also 
seen in Fig.\ \ref{resp-wqe}, where the ratio of the 
$(+,+)$--projected (i.e., FPWIA) to the full 
$(+,+)\oplus (+,-)\oplus (-,-)$ (i.e., RPWIA) 
responses is shown, for quasifree kinematics.
The $\lambda=1$ vertex functions are used, in order
to display the maximum effect. The $R_T$ response 
obtained with the $\Gamma^\mu_1$ EM vertex 
shows $>2$\% effects below $\bq=600$ MeV/c. 
At the SAMPLE kinematics ($\bq=300$ MeV/c)
the effect on $R_T^{cc1}$ is 5\%.

Returning to Fig.\ \ref{pr1_1000}, 
note that the $(-,+)$ terms are 
not always dominant over the  $(-,-)$ ones
(see, e.g., the $\lambda=1$ panels), 
as would have been  expected given 
the extra suppression of the latter by $\fm/\fp$. 
Finally, the $\lambda=1$ results
in Fig.\ \ref{pr1_1000} show an observable, 
but not dramatic change when the EM form factors 
are modified according to \equ(mod_naus).

\begin{figure}[htb]
\begin{center}
 \mbox{\epsfig{file=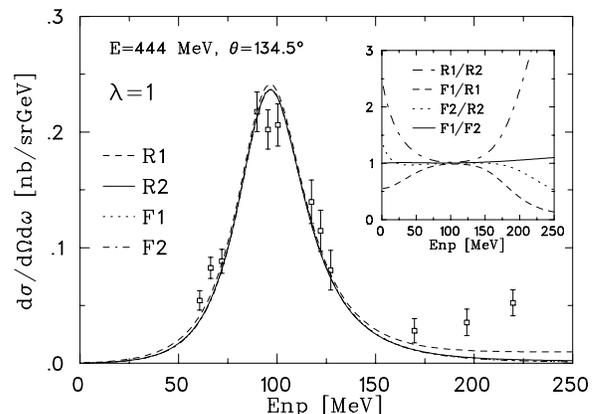,width=0.45\textwidth,angle=90}}
\caption[dummy]{Inclusive $d(e,e')$ cross section using the covariant
               ($R_i$) and $(+,+)$--projected ($F_i$) responses,
               as a function of the final state excitation energy
               $E_{np} = \sqrt{(\md+w)^2-\bq^2}-2m$. The
                $\lambda=1$ vertex functions and both
                $i=1,2$ EM vertex prescriptions are used.
                Ratios between different calculations are 
               displayed in the inner graph. 
               Data from Ref.\ \cite{s444}.}
\label{sigma444}
\end{center}
\end{figure}

Turning to a direct comparison between cc1 
and cc2 results, we write, using \equ(ccvertex),
\be\label{cc1cc2}
\bar u(p)\left[\Gamma^\m_2 - \Gamma^\m_1\right] 
        = {F_2\over 2m} \bar u(p)\left[ 
\left(\Delta^\m - \gamma^\mu\s\Delta\ri) 
+
\g^\m\left(\s\bk - m\ri) \ri]\ ,
\ee
where we have set $\Delta=\bk-k=(\E-\Ek,{\bf 0})$.
The $\Delta$ terms do not induce a large effect.
In fact, they do not contribute anything
when $\mu=0$. In the FPWIA, the  $(\s\bk-m)$ term does not contribute
because of the $(\s\bk+m)$ projector onto the $(+,+)$ sector.
Thus, the cc1 and cc2 longitudinal responses (both $R^L$
and its parity-violating counterpart, $R^L_{AV}$) are
{\em identical} in FPWIA~\cite{thesis,Juan} while the transverse
ones differ, though not by much. This is not discernable 
from Figs. \ref{pr2_300}--\ref{resp-wqe}, but can be seen 
from Fig. \ref{sigma444}, especially the inner plot, where
the FPWIA cross sections differ insignificantly between cc1 
and cc2 throughout the excitation energy region (solid line).
The $\left(\s\bk - m\ri)$ term, however, contributes to the 
$(+,-)$ and $(-,-)$ parts of the RPWIA. It is therefore not
surprising that a significant difference is seen in Fig.\ 
\ref{sigma444}  between the cc1 and cc2 cross sections in RPWIA
(dot-dashed line in inner plot) for non-quasifree kinematics, to be
contrasted with the coresponding behavior after $(+,+)$ 
projection (FPWIA, solid line).

Notice also, from the $\sigma^{++}_{\rm cc1}/\sigma_{\rm cc1}$
(labelled $F_1/R_1$) and $\sigma^{++}_{\rm cc2}/\sigma_{\rm cc2}$ 
(labelled $F_2/R_2$) ratios in the
inner graph of Fig.\ \ref{sigma444},  
that the negative energy components play a larger 
role when the $\Gamma^\m_1$ vertex is used.
In fact, this is a common and rather unexpectred 
feature of all Figs. \ref{pr2_300}--\ref{sigma444}:
{\em the effect of the projection is more significant
when cc1 is employed}. Given that
either EM vertex amounts to an equally {\it ad hoc} choice, 
this observation cannot justify the overwhelmingly dominant 
use of the `cc1' prescription in data analyses,
if not suggesting that relativistic effects are more
appropriately incorporated when using cc2.

\subsection{Parity-violating responses and asymmetry}

In inclusive $(\vec e, e')$ scattering of longitudinally 
polarized electrons (helicity $h=\pm 1$) from an 
unpolarized target, the asymmetry  $A$ = 
$ (d\si^+ - d\si^-)/(d\si^+ + d\si^-) $ is 
purely parity-violating. There are two sources
of parity-violation, electroweak ($\gamma-Z^0$) 
interference, and nuclear parity-violation.
In the latter case, the weak interaction leads to a
parity-violating component in the NN Hamiltonian, which
allows parity mixture with $P$ states (not to be confussed
with the relativistic $P$ states discussed so far, which
correspond to lower components of the coupled Dirac 
wavefunctions and therefore have the same {\it overall} 
parity as the $S$ and $D$ upper components~\cite{BGross}; 
these $P$ states, on the other hand, arise in a nonrelativistic
framework and have genuinely opposite parity). 
Nuclear parity-violation appears 
suppressed relatively to $\gamma-Z^0$ interference
by $\alpha/\alpha_s$, where $\alpha_s$
denotes the strength of
the (parity-conserving) strong interaction. In a recent
calculation~\cite{Gun} the nuclear PV asymmetry appears, indeed, 
negligible (suppressed by three orders of magnitude)
compared to the asymmetry due to electroweak interference
over the whole range of momentum transfer values 
considered ($ {\bf q}\in [0,1000] $ MeV/c).
Thus, in the following we focus on the PV asymmetry generated
from the interference of electromagnetic (photon-exchange, $A_\g$)
 and neutral current ($Z^0$-exchange, $A_{Z}$) amplitudes.
In the one-boson approximation,  and for momentum 
transfer values much smaller than the electroweak
scale, $|q^2|<<M_W^2$, the asymmetry is  given by~\cite{rev}
\be\label{asym1}
A = {G |q^2|\over 2\pi\alpha\sqrt{2}}{a_A\left[ v_L R^L_{AV} + v_T R^T_{AV}\right]
                  + a_V v_{T'} R^{T'}_{VA}
                \over v_L R_L + v_T R_T  }     \ .
\ee
Here $G$ is the Fermi constant and $a_V$($a_A$) are the 
vector(axial-vector) couplings in the $e-Z^0$ vertex
($a_V \gamma_\m + a_A\g_5\g_\m$). To understand the 
$V-A$ structure of \equ(asym1), it is useful to observe that the 
helicity ($h=\pm 1$) dependent part of the 
leptonic tensor $\tilde l^{\m\n}$ 
corresponding to the $A_\g - A_Z$ interference, reads 
\bea\label{lep_nc}
\tilde l_{\m\n} &\equiv& \sum_{h'}
      <e',h'|a_V \gamma_\m + a_A\g_5\g_\m|e,h>^* 
<e',h'|\g_\n|e,h> \nnn\\
               &\sim& h(a_A s_{\m\n} + i a_V a_{\m\n}) + 
             \left[\mbox{$h$--independent}\right] \ ,
\eea 
where $s_{\m\n}(a_{\m\n})$ 
are the symmetric(antisymmetric) parts of the leptonic tensor 
in \equ(lep_EM), corresponding to the purely EM 
photon-exchange amplitude squared, $|A_\gamma|^2$~\cite{thesis}. 
Analogously, we decompose the hadronic 
tensor corresponding to the $A_\gamma-A_Z$ interference,
\be\label{had_nc}
\tilde H^{\m\n}\equiv \b1sum_{\rm spins}
 <f|\hat J^\m_{NC}|i>^* <f|\hat J^\n_{EM}|i> \ ,
\ee
into $\tilde H^{\m\n}=S^{\m\n}+iA^{\m\n}$. For unpolarized 
targets, an antisymmetric part $A^{\m\n}$
may only arise from axial-vector coupling. From general 
considerations in inclusive electron scaterring~\cite{Don_Rask} 
with polarized beam and unpolarized target,
the $L$ and $T$ responses are generated by contracting the
symmetric parts of the leptonic and hadronic tensors. Thus, from
\equ(lep_nc) the  $L$ and $T$ parity-violating responses in the
numerator of \equ(asym1) arise from interference of the leptonic 
axial-vector with hadronic vector pieces (hence the $AV$ 
index). The $T'$ response, on the other hand, 
arises from contracting the antisymmetric tensors; 
it does not, therefore, contribute in unpolarized 
scattering -- hence, does not enter the helicity sum in 
the denominator of \equ(asym1) --  and has a leptonic/vector
 with hadronic/axial-vector ($VA$) interference character.
Thus, the nucleon axial-vector form factor $\tG_A^{T=1}$ that 
plays an important role in motivating these studies, 
as described in Sect.~I, appears in the 
$R^{T'}_{VA}$ response. Notice, though, that since in 
the Standard Model $a_V = -1 + 4\sin^2\theta_W 
\simeq -0.092$, while  $a_A = -1$, the $R^{T'}_{VA}$ response
is effectively suppressed with respect to the other 
responses in electron scattering (but not in neutrino 
scattering, where $a_V=1$~\cite{rev}).

The PV responses are computed from an interference hadronic
tensor $\tilde H^{\mu\nu}$, analogous to the 
purely EM one in \equ(proton2), 
\be\label{nc_tensor}
\tilde H^{\mu\nu}_{\rm RPWIA}
 =  {4m^2\over 3(k^2\!-\!m^2)^2}\  {\rm Tr}\left\{
	{\s S + T\over 2m}\ \bar \Gamma^\m_{NC}\ {\s p + m\over 2m}\ 
	\Gamma^\n_{EM} \right\} \ ,
\ee
and then integrated according to \equ(s1).
In a system of axes where the momentum transfer defines the 
z-axis and the y-axis is taken perpendicular to the 
hadronic plane defined by $\v q$ and $\v k$,
one has $W^T_{AV} = \tilde H^{11}+\tilde H^{22}$, $W^{T'}_{VA} = 
-2 \mbox{Im} \tilde H^{12}$, and, in the ``cc'' treatment 
of current conservation, $W^L_{AV} = \tilde H^{00}$.
In RPWIA we use, in direct analogy with \equ(ccvertex),
the momentum-space vertex operators
\bea\label{g2_nc}
\Gamma^\mu_{NC, 2} &=& \tilde F_1\gamma^\mu + i{q_\tau\over 2m}
	\tilde F_2\sigma^{\mu\tau} 
	+ \tilde G_A\gamma_5\gamma^\mu \nnn\\ 
\Gamma^\mu_{NC, 1} &=& \left(\tilde F_1 + \tilde F_2\right)\gamma^\mu 
	- {\left(k+p\right)^\m \over 2m}\tilde F_2 
	+ \tilde G_A \gamma^\m \gamma_5 \ ,
\eea
where we have not included terms corresponding to
second-class currents~\cite{rev}. An induced pseudoscalar
axial-vector term $\tG_P q^\m\g_5$ is possible and will
generate a $\tG_P q^\m[\dots]^\n$ term in 
the hadronic tensor. Since axial contributions appear only 
in $W^{T'}_{VA} = -2 \mbox{Im} \tilde H^{12}$, though, 
and $q^\m$ has vanishing 1 and 2 components,
this term does not contribute.

\begin{figure}[htb]
\begin{center}
 \mbox{\epsfig{file=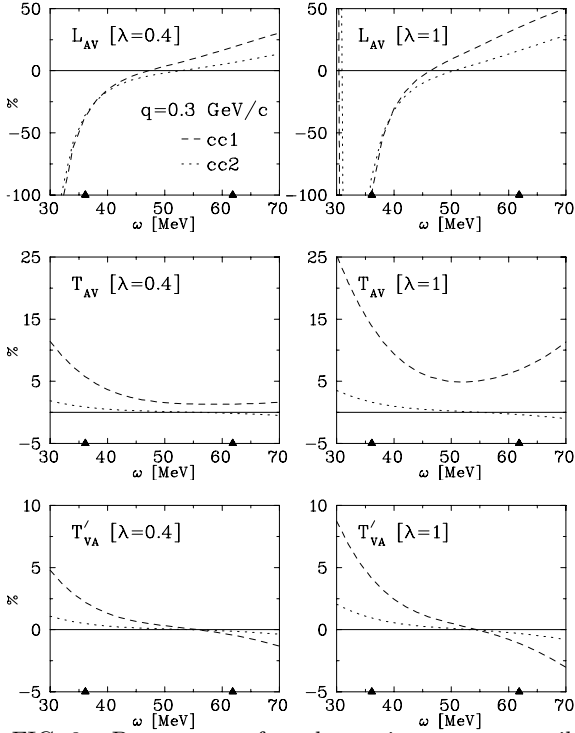,width=0.45\textwidth,angle=0}}
\caption[dummy]{ Percentages of total negative-energy contribution
              ($1-R^{++}/R$) to the parity-violating 
                ($L_{AV},T_{AV},T'_{VA}$) responses
                 computed in FPWIA ($R^{++}$) and RPWIA $(R)$
                  with both cc1 (dashes) and cc2 (dots)
                 at $\bq=300$ MeV/c. The $\lambda=0.4(1)$ 
                 vertex functions are used in the left(right) panels.
                 Notice the difference in scales.
        }
\label{weak-300}
\end{center}
\end{figure}

In the Standard Model, the electromagnetic (EM) and 
weak neutral (NC) current operators (and therefore the corresponding
nucleon form factors), are written as different linear 
combinations of elementary vector, $\bar q\gamma^\m q$,
and axial-vector, $\bar q\gamma^\m\gamma_5 q$  (NC only),
currents. Thus, considering three flavors ($q=u,d,s$) only, 
and using isospin symmetry, 
the NC form factors (in Sachs, $\tG_E = \tF_1 - 
\t \tF_2$, $\tG_M = \tF_1 + \tF_2$, $\t=|q^2|/4m^2$, representation) 
can be cast in the form~\cite{rev,Hadj}
\bea\label{ff_nc}
\tG_{E,M}^{p,n} &=& \left[  \beta_V^p G_{E,M}^{p,n} 
                          + \beta_V^n G_{E,M}^{n,p} \ri]
                -{1\over 2} G_{E,M}^s \nnn\\
\tG_A^{p,n}     &=&  \beta_A^p G_A^{p,n} + \beta_A^n G_A^{n,p} 
                -{1\over 2} G_A^s \ ,
\eea
where, at tree-level in the Standard Model,
\bea\label{SM}
\beta_V^p &=& {1\over 2}\left(1 - 4\sin^2\theta_W\ri) \nnn\\
\beta_V^n &=& \beta_A^n = - \beta_A^p = -{1\over 2} \ .
\eea
\begin{figure}[htb]
\begin{center}
 \mbox{\epsfig{file=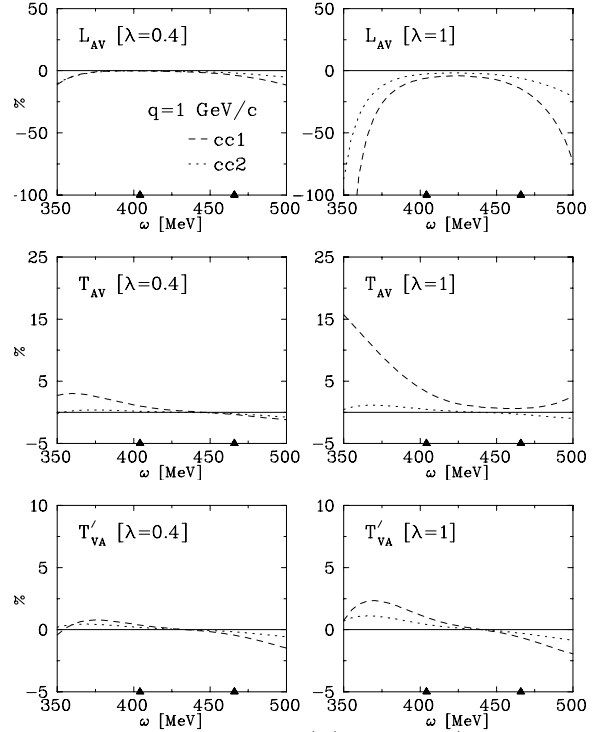,width=0.45\textwidth,angle=0}}
\caption[dummy]{ As in Fig.\ \ref{weak-300}, at $\bq=1$ GeV/c.
                 Notice the difference in scales.}
\label{weak-1000}
\end{center}
\end{figure}

Provided the coefficients $\beta$ are known
(i.e., the radiative corrections small or reliably estimated) 
\equ(ff_nc) implies that independent measurements of the 
EM and NC vector current nucleon matrix elements
allows the experimental extraction of the hitherto
unconstrained strangeness nucleon
form factors $G_{E,M}^s$. The axial-vector
form factors $G_A^{T=1}=G_A^p-G_A^n$, 
$G_A^{T=0}=G_A^p+G_A^n$ and $G_A^s$ will also 
enter the parity-violating asymmetry A, through the
$R^{T'}_{VA}$ response. Of those, $G_A^{T=0}$
does not contribute at tree-level, since, 
from \equ(SM), $\beta_A^{(0)}\equiv \beta_A^p + 
\beta_A^n = 0$. Any dependence on the 
(pressumed small) $G_A^s$ is practically eliminated
due to the aforementioned suppression 
of $R^{T'}_{VA}$ by $a_V/a_A \simeq 10\%$ and has to
be investigated using neutrinos~\cite{rev}.
Thus, only $G_A^{T=1}$ enters, but, as already mentioned, 
the large theoretical uncertainty in the radiative corrections
to the associated coefficient $\beta_A^{(1)}$ 
makes its contribution crucial. Following 
Refs.\ \cite{Hadj,rev}, quantitative predictions are
obtained by generalizing the Galster parametrization
to include the extra form factors appearing in \equ(ff_nc)
\bea\label{gGal}
G_A^{T=1} &=& g_A^{(1)} (1+\lambda_A\tau)^{-2}\nnn\\
G_M^s &=& \mu_s (1+\lambda_s \tau)^{-2}\nnn\\
G_E^s &=& \rho_s \tau (1+\lambda_s \tau)^{-2} \ ,
\eea
where, from $\beta$ decay, at $\t=0$, $g_A^{(1)} = 1.26$, 
and we use $\lambda_A = 3.53$, $\lambda_s = 4.97$, i.e.,
equal to the Galster dipole strength. 
Results in this section are with vanishing strangeness 
magnetic moment, $\mu_s$, and strangeness radius, $\rho_s$~\cite{rev}.

In Figs.\ \ref{weak-300} and \ref{weak-1000}
we show the effects of the projection on the PV responses.
For brevity, we show results only for two 
choices, $\lambda=0.4$ (pv) and $\lambda=1$ (ps), and, moreover,
we do not show individual $(+,-)$ and $(-,-)$ decompositions, rather
only the cumulative effect of the negative energy components. 
The behavior is similar to that of the parity-conserving, 
electromagnetic responses. Specifically, the effects of 
the projection are negligible with pseudovector coupling 
($\lambda=0$, not shown) for all responses and current prescriptions,
and become more important as $\lambda\rightarrow 1$;
in all cases the effects are minimal for quasifree kinematics;
here, as well, the projection affects cc1 more severely
than cc2; inside the quasielastic ridge, the effects 
are more pronounced at low momentum transfer, and as 
with the transverse response at $\bq=300$ MeV/c
in Fig.\ \ref{pr1_300}, a potentially  worrisome 
5\% effect is seen for the quasielastic 
$R^T_{AV}$ cc1 response calculated with pseudoscalar coupling. 
The only essentially new feature comes from the
parity-violating longitudinal response, $R^L_{AV}$, 
which appears extremely sensitive to the projection, 
especially towards threshold, with both  $\lambda=0.4$  
and  $\lambda=1$ vertex functions at $\bq=300$ MeV/c,
and, to a lesser extent, i.e., only with the  $\lambda=1$ ones, 
at $\bq=1$ GeV/c. This sensitivity, however, 
is not a consequence of large dynamical effects associated 
with negative energy components {\it per se}. Rather,
it is an example of {\em any} subdominant effect playing an important
role whenever the leading contribution vanishes (or becomes too small),
as happens in this case because $\sin^2\theta_W\simeq 1/4$. 
To see that, notice from \equ(ff_nc) and the numerical values of the 
coefficients in \equ(SM) ($\beta_V^0\approx 0.046$) 
that, although $G_E^n$ is small compared to $G_E^p$ 
(at least for $\t<1$), in the case of the neutral current 
(NC) it is the {\em neutron} form factor, 
$\tG_E^n$, that is much larger than
the proton one, $\tG_E^p$. 
Consider then, from \equ(trace_cc1) and \equ(reduce),
the cc1 longitudinal response in FPWIA
\bea\label{l_cc1}
\W^{L, cc1}_{VA} &=& \left[ E_n+E_p\over 2 m\ri]^2 \sum_{p,n}{
 \tG_E G_E +  \t \tG_M G_M \over 1+\t }
\nnn\\ &&\qquad -  {\bq^2\over 4m^2} \sum_{p,n} 
\tG_M G_M +{\cal O}(\bk^2-k^2)\ .
\eea
Using \equ(ff_nc) and \equ(SM), and employing the Galster 
parametrization, we have
\bea\label{apli}
{\W^{L, cc1}_{VA}\over 
[1+\t]^{-1}\le( G_E^p\ri)^2} &=& \le[ 1+{\w\over 2m}\ri]^2
\le(\beta_V^p + {\mu_n\t\over 1 + 5.6\t} \ri) \nnn\\
-\t^2 \Bigl(\beta_V^p[\m_p^2 \!&+&\! \m_n^2] - \m_p\m_n\Bigr) 
\le[ 1-{\w\over 2m\t}\ri]^2 
\ .
\eea
where we have left out small contributions from
off-shell, strangeness, and $\t\beta_V^p$ terms.
The first tem in \equ(apli) arises from the electric terms
in \equ(l_cc1), and the second from the magnetic terms.
Because $\beta_V^p$ is small, and the magnetic terms compete
in \equ(l_cc1), there is no dominant 
(i.e., ${\cal O}(1)$) term in the r.h.s. of \equ(apli).
Thus, the parity violating longitudinal response
is {\em small} in magnitude and appears with 
either positive or negative sign, depending on the kinematics.

\begin{figure}[htb]
\begin{center}
 \mbox{\epsfig{file=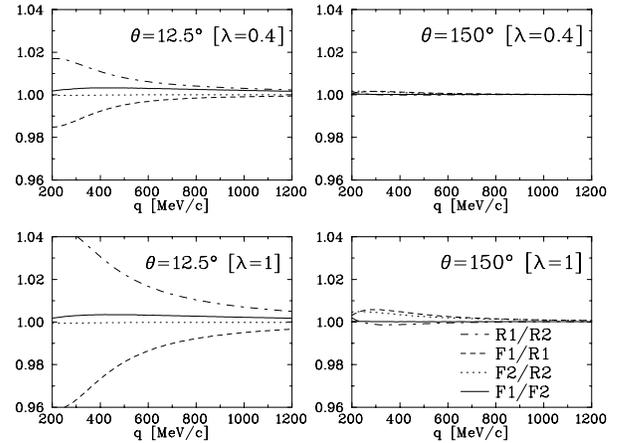,width=0.47\textwidth,angle=90}}
\caption[dummy]{Asymmetry ratios for quasifree kinematics, $\w=\w_{qe}$.
               Reasults presented for two choices of electron scattering
               angles $\theta$ and vertex functions $\lambda$.
               Notation as in Fig.\ \ref{sigma444}.
        }
\label{A-wqe}
\end{center}
\end{figure}

Consider first quasifree kinematics, $\w = q^2/2m$.
In this case the second term in \equ(apli) vanishes 
(there are no magnetic
contributions in the ``static'' approximation~\cite{Hadj}).
From the first term, it is clear that at low $\t$ values 
(and thus low $\bq$, since, for quasifree kinematics, 
$\bq=2m\sqrt{\t(1+\t}$\ ) the quasielastic PV longitudinal response 
is positive and at higher $\tau$ negative, the sign reversing at
$\tau=\beta_V^p/(1.91-5.6\beta_V^p)=0.0278$, 
or $\bq\simeq 317$ MeV/c. The exact kinematics 
where the sign flips will depend, of course, on the 
small effects not included in \equ(apli), as well as the projection
effects [cf. Table.\ \ref{table1}]. 
Thus, around $\bq=320$ MeV/c the zeros in  $R^L_{AV}$
give rise to a discontinuous behavior of the 
the ratio of $R^L_{AV}$ in 
FPWIA and RPWIA for quasifree kinematics, as indeed 
observed in Fig.\ \ref{resp-wqe}. Interestingly enough,
this momentum transfer value is very close to that of
SAMPLE. SAMPLE, however, focusing on the magnetic 
strangeness form factor, is a backward angle measurement.
At $\theta > 50^o$ the asymmetry is dominated by
the $T$ part, and, to a lesser extent -- because of 
$a_V$ being small -- by the $T'$ part. Thus, 
the PV longitudinal response, and the projection effects
associated with it, do not affect the SAMPLE measurement.

Because of the behavior of $R^L_{AV}$, one expects 
that projection effects will depend strongly on 
the electron scattering angle.
For quasifree kinematics, this is evident 
from Fig.\ \ref{A-wqe}.
At $\bq=300$ MeV/c, up to 4\% effects in the forward
asymmetry can be seen (this would be prohibitively large for
strangeness studies), while at backward angles the asymmetry 
is very stable, showing a maximum in the $\bq=300$ MeV/c region, 
which is, nevertheless, less than 1\%. Recall that, in this 
case, the individulal $R^T$ and $R^T_{AV}$
transverse responses appearing, respectively, in the numerator
and denominator of the asymmetry, show a ${\cal O}(5\%)$
projection effect with cc1 and $\lambda=1$ [cf. Figs.\ 
\ref{pr1_300}, \ref{weak-300} and \ref{resp-wqe}], 
even on the quasielastic peak. 
Due to the dominantly transverse character of the asymmetry
at $\theta=150^o$, however, these effects by-and-large 
cancel in the ratio, and at this angle the asymmetry 
shows the aforementioned $<1$\% variation.
The extent to which such an effect may interfere with strangeness
studies is examined in the next section. Fortunately,
the projection effects become less than 1\% for high enough 
momentum transfer even at forward angles, 
as, e.g., for a possible TJNAF Hall A measurement 
($\bq = 1$ GeV/c, $\theta=12.5^o$) proposed in Ref.\ \cite{Hadj}.

\begin{figure}[htb]
\begin{center}
 \mbox{\epsfig{file=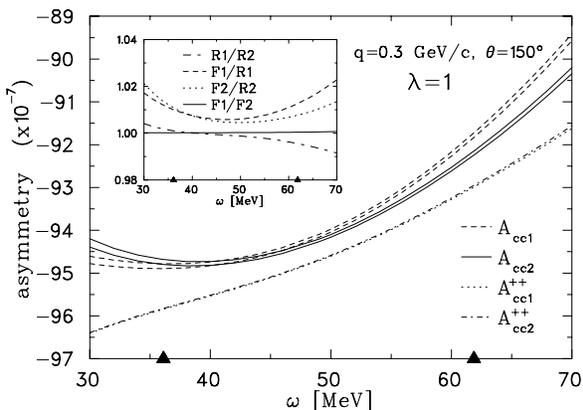,width=0.45\textwidth,angle=90}}
\caption[dummy]{ The PV asymmetry at $\bq=300$ MeV/c and 
                backward  ($\theta=150^o$) angle, using the
                  $\lambda=1$ Buck-Gross vertex functions.
               RPWIA results are shown 
               with (lower) and without (upper curves)
               the modification of the on-shell form factors
               according to \equ(mod_naus).
               Notation in inner graph as in Fig.\ \ref{sigma444}.
        }
\label{A-300}
\end{center}
\end{figure}

Away from the quasielastic peak, the second term in
\equ(apli) gives a negative definite contribution, with
a coefficient that rises with $q^2$. Thus, 
$R^L_{AV}$ is still positive for low $\bq$
(as e.g., at 300 MeV/c) and negative at  large $\bq$
(as e.g., at 1 GeV/c). It is always small, though, and,
especially outside the quasielastic ridge, negative 
components constitute a significant part of the response,
resulting in the behavior observed in Figs.\ \ref{weak-300}
and \ref{weak-1000}. At such non-quasifree kinematics, however,
$R^L_{AV}$ is quite sensitive to other aspects of 
nuclear dynamics beyond PWIA, as well, as has been 
shown in studies of both deuterium
(e.g., proton-neutron interference contributions~\cite{pwba}) 
and heavier nuclei~\cite{Don_RL}.

\begin{figure}[htb]
\begin{center}
 \mbox{\epsfig{file=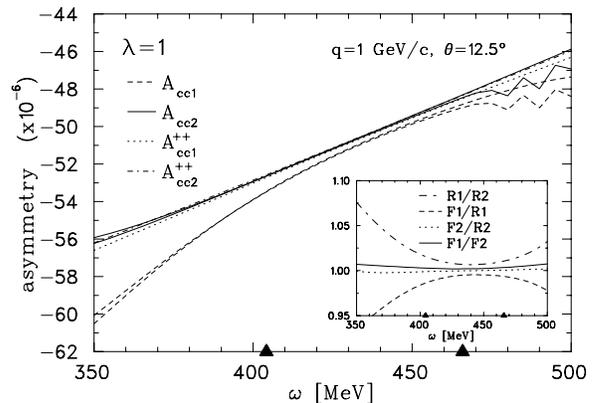,width=0.45\textwidth,angle=90}}
\caption[dummy]{ As in Fig.\ \ref{A-300}, at $\bq=1$ GeV/c and forward 
          ($\theta=12.5^o$) angle.
        }
\label{A-1000}
\end{center}
\end{figure}

In Figs.\ \ref{A-300} and \ref{A-1000} we show the
asymmetry in FPWIA and RPWIA, as well as the corresponding 
ratios, for the SAMPLE kinematics ($\bq =300$ MeV/c, backward angle)
and a possible $\bq=1$ GeV/c forward angle measurement at TJNAF. 
The $\lambda=1$ vertex functions are used to display
the maximum effect of the projection. The residual relativistic 
effects associated with the relativistic treatment 
of the deuteron and the struck nucleon induce a less 
than 2\% variation of the PV asymmetry within the 
quasielastic ridge. We also observe that 
the modification of the on-shell form 
factors to allow for a $k^2$-dependence according to \equ(mod_naus)
amounts to a secondary effect (i.e., much smaller than the
effect caused by the projection). An exception is observed
at high $\w$ values in the $\bq=1$ GeV/c case. This behavior is 
caused when deeply off-shell, $k^2={\cal O}(-m^2)$, momenta,
for which \equ(mod_naus) generates unreasonably large modification 
factors, contribute to the inclusive integral in \equ(s1). 
Since $k^2 = (\md-\E)^2-\k^2 \simeq 5m^2-4m\sqrt{m^2+\k^2})$, 
values $k^2 = -m^2$ occur in the inclusive integral 
for $\k\ge 1.1 m$. Such values are accessible for 
$\bq>m$  and with increasing $\w$ values, since above 
the quasielastic peak ($y>0$) the upper integration 
limit $Y=y+\bq>\bq$. 

Another qualitative difference between 
the two cases, is that while at $\bq=300$ MeV/c 
the behavior of cc1 is quite similar to that of cc2 on the quasielastic peak 
for both FPWIA (solid lines in inner graph of Fig.\ \ref{A-300})
and RPWIA (dot-dashes), at  $\bq=1$ GeV/c 
we see (as we observed for the cross section
before) that projection effects are larger for cc1 than cc2.
At $\bq=300$ MeV/c the magnitude of the asymmetry is raised
in going from RPWIA to FPWIA by about $0.5-1.5$\% with either 
cc1 or cc2 (Fig.\ \ref{A-300}), while at $\bq=1$ GeV/c 
(Fig.\ \ref{A-1000}) there is an observable effect only for cc1, 
the magnitude of the asymmetry being reduced by $0.5-1$\%
inside the quasielastic ridge.

\subsection{Strangeness studies}

\begin{figure}[htb]
\begin{center}
 \mbox{\epsfig{file=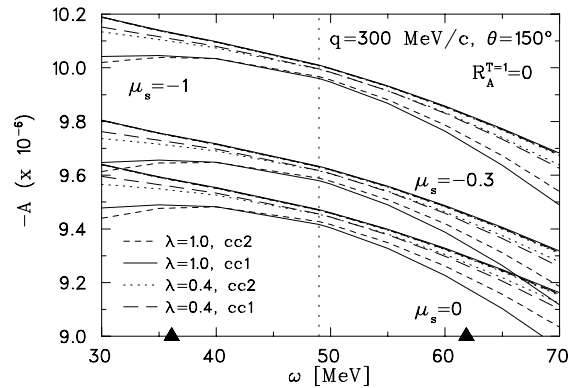,width=0.45\textwidth,angle=90}}
\caption[dummy]{ The PV asymmetry at $\bq=300$ MeV/c and
                 $\theta=150^o$.  Results shown for 3 values of
                 the strangeness magnetic moment  $\mu_s$ 
                 and 2 choices of Buck-Gross vertex functions 
                 $\lambda$. For each ($\mu_s,\lambda$) choice, results
                 are shown with both cc1  and cc2 for RPWIA (lower) and
                 FPWIA (upper, practically indistinguishable curves). 
                 The quasielastic  peak is marked by the dotted 
                 vertical line.
        }
\label{sample}
\end{center}
\end{figure}

Here we discuss the extent to which
the residual relativistic effects associated with the
projection affect the extraction of strangeness
form factors from the PV $e-d$ asymmetry. We are primarily
interested in the interplay between the magnetic 
strangeness form factor $G_M^s$ 
and the NC axial isovector form factor $\tG_A^{T=1}$
(more precisely, the radiative corrections
$R_A^{T=1}$ to the relevent coefficient
$\beta_A^{(1)}=1.26[1+R_A^{T=1}]$). The 
approved SAMPLE and E91-017 experiments will attempt to extract
both quantities by combining measurements of 
the $e-p$ and $e-d$ PV backward-angle asymmetries.
 In Fig.\ \ref{sample} we show 
``bands'' of RPWIA and FPWIA results for SAMPLE kinematics,
with $R_A^{T=1}=0$ and for three values 
of the magnetic strangeness moment $\mu_s$ entering the 
{\it Ansatz} in \equ(gGal), corresponding to zero strangeness ($\mu_s=0$), 
the pole model result of Jaffe ($\mu_s=-0.3$)~\cite{strange} 
and an unlikely high value ($\mu_s=-1$). Within each $\mu_s$
``band'', the projected (FPWIA) results show negligible variation 
(0.03\% on the peak) between different calculations (i.e.,
cc1 {\it vs.} cc2 and $\lambda=1$ {\it vs.} $\lambda=0.4$). 
The variation induced
when $P$ states and negative energy components are included
is small, but not negligible. Although the $\mu_s=1$ and $\mu_s=0$
bands are well separated, outside the quasielastic ridge
the $\mu_s=0$ and $\mu_s=-0.3$ bands overlap. As hoped, though,
they have a minimum width for quasifree kinematics. This 
width depends, of course, on $\lambda$, as is evident 
from Fig.\ \ref{sample}: the $\lambda=0.4$ (long-dashes)
cc1 RPWIA curves are much closer to the overlapping upper 
FPWIA curves compared to the $\lambda=1$ (solid) ones. 
On the quasielastic peak, the maximum variation is 
observed between the projected and unprojected $\lambda=1$ 
cc1 curves, amounting to a 0.56\% effect. 
The asymmetry is linear in $\mu_s$, $A(\mu_s) = 
A(0)[1-b_\m\mu_s]$, with $b_\m=0.057(1)$ in our models.
Although $b_\m$ is approximately tenfold the 
sensitivity to projection effects, 
it may not be large enough if $\mu_s$ turns out to
be rather smaller than the pole value (something not excluded from other 
calculations~\cite{strange}), to the extent, of course,
that the widths observed in Fig.\ \ref{sample} 
give a measure of the theoretical nuclear physics 
uncertainty in describing the $e-d$ reaction.

\begin{figure}[htb]
\begin{center}
 \mbox{\epsfig{file=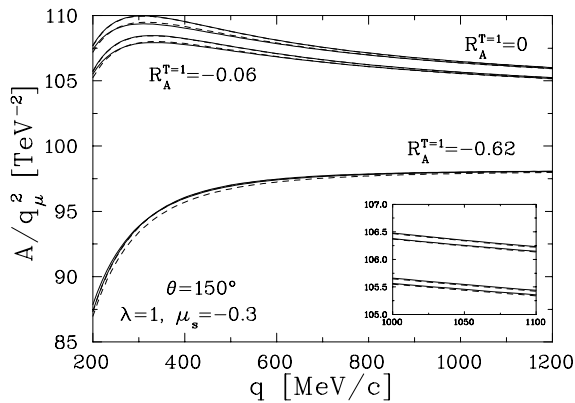,width=0.45\textwidth,angle=90}}
\caption[dummy]{ The PV asymmetry [in units of $q^2$] as function of
                 momentum tranfer $\bq$ for quasifree kinematics
                 and $\theta=150^o$.  Results shown for 3 values of
                 $\beta_A^{(1)}$ (see text). Solid lines: cc1, dashes:
                 cc2. A magnification of the 1 GeV/c region is shown in the
                inner graph.
        }
\label{ga}
\end{center}
\end{figure}

In view of the rather small sensitivity of the $e-d$ asymmetry
to $G_E^s$, the current consensus is to use the deuteron as
a means of constraining the large theoretical uncertainty in 
$R_A^{T=1}$, using primarily the proton for extracting
$G_M^s$. Accordingly, in Fig.\ \ref{ga} we fix the strangeness 
magnetic moment to the pole model value, and show ``bands'' 
of quasifree kinematics results,  
at tree level, $R_A^{T=1}=0$, as well as two other values 
corresponding to the extrema in the calculation of 
Ref.\ \cite{rat1}, $R_A^{T=1}=-0.34\pm 0.28$~\cite{SAMPLE}. It is evident that
significant constraints can be placed on $R_A^{T=1}$ from a measurement
of the asymmetry and that the residual relativistic effects discussed
here should not inhibit this procedure. Comparing the 
$\bq=300$ MeV/c and $\bq=1$ GeV/c regions,  however, it appears that
a higher $q^2$ measurement than that of SAMPLE (as for example 
in the E91-017 TJNAF experiment~\cite{TJNAF}) will place tighter constraints
on $R_A^{T=1}$.   Specifically, since the asymmetry is linear
in $\tG_A$, we may write $A(\mu_s,R_A^{T=1})=A(\mu_s,0)[1
+b_A R_A^{T=1}]$. At $\bq=300$ MeV/c, the typical 
variation between models is $\Delta_{th}=0.56\%$, 
while the sensitivity to the radiative corrections
$b_A=0.235(4)$. On the other hand, at $\bq=1050$ MeV/c
we find a smaller $b_A=0.1259(1)$, but also considerably reduced
model variation, $\Delta_{th}=0.1$\%. From the $A(\mu_s,0)$
values we find that the difference between the  $R_A^{T=1}=0$
and  $R_A^{T=1}=-0.06$ bands (upper two ones in Fig.\ \ref{ga}) is 
$2.4\Delta_{th}$ at $\bq=300$ MeV/c, but roughly 
three times more effective, $7\Delta_{th}$, 
at $\bq=300$ MeV/c.


\begin{table}[htb]
\renewcommand{\arraystretch}{1.3}
        \caption[dummy]{ Momentum transfer $|{\bf q}_{cr}|$ in MeV/c
                  where $R^L_{AV}$ becomes negative (quasifree
                  kinematics, $\mu_s=-0.3$). Results
                  shown using: the static estimate in \equ(apli2),
                  factorized (F) cc1, and covariant (R) 
                  PWIA cc1 with the $\lambda=1$ vertex functions.
                }
        \protect\label{table1}
\begin{tabular}{cccccccc}
        \hline
     $\rho_s$
     & $1.5$
     & $0.5$
     & $0$
     & $-0.5$
     & $-1.0$
     & $-1.2$
     & $<-1.4$
           \\    
        \hline
$R_{cc1}$  & 259 & 295 & 321  & 360  & 430 & 493 &  $-$\\ \hline
$F_{cc1}$  & 258 & 294 & 320  & 357  & 422 & 476 &  $-$ \\ \hline
\equ(apli2)& 257 & 292 & 317  & 354  & 417 & 465 &  $-$ \\ \hline
\end{tabular}
\end{table}

In the previous section we discussed the sign-reversing 
behavior of the PV longitudinal response $R^L_{AV}$ in 
the absence of strangeness. For quasifree kinematics, where
the magnetic terms cancel, only $G_E^s$ enters, and
a negative value of the strangeness radius modifies significantly 
this behavior. Using \equ(gGal), \equ(apli) 
at $\w=2 m\t$ now becomes
\bea\label{apli2}
\W^{L, cc1}_{VA}  \sim  
\le(\beta_V^p + {\m_n\t\over 1 + 5.6\t} -
\t{\rho_s\over 2}\left[1-{\m_n\t\over 1 + 5.6\t} \right]
\ri) \ .
\eea

The resulting quadratic equation gives an estimate
of the momentum transfer value $|{\bf q}_{cr}|$ 
where $R^L_{AV}$ reverses sign as a function of $\rho_s$. 
As seen from Table\ \ref{table1}, this ``static''
estimate is reasonably accurate. For positive strangeness 
radius $|{\bf q}_{cr}|$ depends mildly on $\rho_s$, but
for negative values $|{\bf q}_{cr}|$ becomes significantly
larger and for $\rho_s<-1.4$ (the pole model value is $-2.1\pm 1.0$)
the response is always positive.

Thus, given the large variation
in the model predictions of $\rho_s$~\cite{strange},
a determination of $R^L_{AV}$ in different $\bq$ values
might be helpful in constraining $\rho_s$. Although quite 
challenging, a separation of the $L$, $T$ and $T'$
parts of the PV asymmetry is not beyond current experimental
capabilities, and will be performed for the $e-p$ PV asymmetry 
as part of the E91-017 TJNAF experiment~\cite{rev}.

\section{summary}

In this work parity-violating,  electron-deuteron
scattering in the quasifree region has been studied 
in relativistic PWIA. The emphasis has been to 
quantify the role of residual
relativistic effects associated with the non-relativistic
treatment of the bound state in the semirelativistic,
factorized PWIA, assuming that the latter correctly incorporates
relativistic effects associated with the large momentum transfer.
The main results of this investigation 
can be summarized as follows:
\begin{itemize}
\item
     In the covariant calculation (RPWIA), factorization
     is spoiled by off-shell contributions associated
     with the struck nucleon. 
     A weak factorization still holds, with the 
     momentum distribution and the single-nucleon tensor
     being now density matrices in  spin and energy-sign 
     spaces of the struck nucleon. The familiar FPWIA
     using de Forest's single-nucleon tensor corresponds
     to a projection of the RPWIA to the positive energy sector.
     That is, the $H^{\m\n}_{++}$ single-nucleon tensor is the FPWIA
     single-nucleon tensor, and the $\rho_{++}$  momentum distribution
     is the nonrelativistic deuteron momentum distribution. 
     No additional operations (such as neglect of $P$ states that
     superficially appear in $\rho_{++}$, or nonrelativistic expansion)
     is required. Thus, this truncation defines the residual 
     relativistic effects we are interested in, and shows the close 
     connection between relativistic and off-shell effects. 
     Similar considerations apply to quasielastic scattering on
     any nucleus and not just deuterium. 
\item
     Since, from the above, a covariant treatment of the $e-d$
     scattering  requires inclusion of off-shell effects, it is 
     not consistent to incorporate such effects in the
     d-n-p vertex but ignore them in the $\gamma NN$ and $Z^0 NN$ 
     vertices. A measure of the thusly induced model dependence
     is obtained by using various on-shell-equivalent prescriptions 
     for these vertices, and several model calculations for the 
     deuteron vertex functions.
\item
      The residual relativistic effects depend strongly on the
      $\pi NN$ coupling used for calculating the deuteron vertex
      functions (minimal for pseudovector and maximal for 
      pseudoscalar coupling). They also depend on the off-shell
      prescription for the nucleon vertex. Although the difference 
      between cc1 and cc2 prescriptions is negligible in FPWIA,
      it is not such in RPWIA, where projection effects appear
      more pronounced with the cc1 prescription. In that 
      respect, the cc2 presription seems preferable. 
\item
      Effects associated with allowing for an off-shell dependence of
      the nucleon form factors are  generically  smaller
      than projection effects. The arbitrariness of the
      {\it Ansatz} used, however, does not allow definitive
      conclusions to be drawn.
\item  
      With the exception of the parity-violating longitudinal
       response, $R^L_{AV}$, inside the quasielastic ridge 
      the residual relativistic effects are below 5-10\%
      with cc1, and a few \% with cc2, when the pseudoscalar coupling
      (which maximizes such effects) is used. On the quasielastic 
      peak, effects are typically below 2\%. A notable exception
       is the $\bq< 400$ MeV/c region where the transverse responses, 
      $R^T$ and $R^T_{AV}$, can have $>4$\% effects. 
      The $R^L_{AV}$ response can show quite important effects, 
      especially in the $\bq=300$ MeV/c region, which reflect 
      the lack of a dominant term in this response, thereby
      exposing otherwise non-leading contributions. 
\item
      The magnitude of these corrections for quasifree kinematics
      is small for them to play a role in the cross section.
      In the parity-violating asymmetry, where 1\% effects are important
      for strangeness studies, the effects mostly cancel in 
      the ratio. The $R^L_{AV}$
      response is sensitive to these effects and, thus,
      effects $>2\%$ are seen in the forward angle asymmetry
      for $\bq<800$ MeV/c. For backward scattering, 
      where the deuteron measurements will be used for 
      disentangling the correlation between magnetic strangeness 
      and axial-vector isoscalar radiative corrections, 
      the PV asymmetry is sensitive to    
      less than 1\% to residual relativistic effects, the 
      maximum sensitivity oobserved in the $\bq=300$ MeV/c region.
      Thus, the extraction of $\tG_A^{T=1}$ at the E91-017 TJNAF kinematics
      will be less sensitive to such effects than at the SAMPLE kinematics.
\end{itemize}

\acknowledgements

I would like to thank Prof. J.A. Tjon for 
discussions that led to the formalism in Sect.~III,
Dr. A. Korchin for suggesting this calculation,
and Dr. H.W.L. Naus for explaining certain features
of the off-shell EM form factor. This work has been 
supported by Human Capital and Mobility Fellowship
ERBCHBICT941430 and the Research Council of Australia.



\appendix

\section{}

Here we present some details of the calculation of
the hadronic tensor in RPWIA. After defining projected vectors
\bea\label{proj1}
\D^\m &\equiv& \q^\m - d^\n{(d\c \q)\over \md^2} \nnn\\
N^\m &\equiv&  n^\m - d^\n{(d\c n)\over \md^2} \ ,
\eea
a straightforward calculation gives for the scalar term $\beta$
in \equ(defb) 
\bea\label{scalar}
\beta  &=& 3m A^2 + {B^2\over m} \D^2
		-2 {A B\over m}(\D\c n) \nnn\\
	&+& AF\left[ 3m - 3{(k\c n)\over m} 
          + 2{ (k\c N)\over m}\right]\nnn\\ 
	&+& {BG\over m}\D^2\left[1+{(k\c n)\over m^2} 	\right]\nnn\\
	&-& {AG + BF\over m}
	\left[(k\c \D)+(n\c \D)\right]\nnn\\
	&+& {G^2\over 4m^3}\D^2\left[ (m^2+k^2) + 2(k\c n)\right]\nnn\\
	&+& {F^2\over 4 m} \left[ 4(k\c N) - 6(k\c n) + 3(m^2+k^2)\right]\nnn\\
	&+& {FG\over 4m^3}\left[-2(m^2+k^2)(\D\c n)-4m^2(k\c\D)	\right] \ \ ,
\eea
and for the vector term
\bea\label{vector}
\U^\m   &=& A^2\left[3 n^\m -2 N^\m\right]
		- {B^2\D^2\over m^2} n^\m
		+ 2A B\D^\m\nnn\\
	&+& AF\left[ 3n^\m-2N^\m-3k^\m	\right]\nnn\\
	&-& {BG\over m^2}\D^2 \left[ k^\m+n^\m\right]\nnn\\
	&+& {AG+BF\over m^2} \left[m^2\D^\m + (n\c\D)k^\m\right] \nnn\\
	&+&	{AG-BF\over m^2} 
	\left[  (k\c n)\D^\m-(k\c \D) n^\m\right]\nnn\\
	&+& {G^2D^2\over 4m^4}\left[ (k^2-m^2)n^\m -2
	\left[(k\c n)+m^2\right]k^\m\right]
		\nnn\\
	&+& {F^2\over 4 m^2} \left[ \left[6(k\c n)
               -4(k\c N)-6m^2\right]k^\n\right. \nnn\\ 
	& &\qquad\qquad 
           +\left.(k^2-m^2)\left[2 N^\n - 3n^\n\right]\right]\nnn\\
 	&+& {FG\over 4 m^3} \left[ 4m(\D\c n)k^\m  +
             4m(k\c \D)k^\m\right.\nnn\\
	&&\qquad\qquad \left.+2m(m^2-k^2)\D^\m\right] \ .
\eea
Using the $\tilde \Gamma^\m_2$ vertices of 
\equ(g2_nc) and \equ(ccvertex),
the cc2 interference hadronic tensor [cf. \equ(had_nc)] reads
\bea\label{trace_cc2}
&&Z\; \tilde H^{\mu\nu}_{cc2} =
\tilde F_1 F_1\left[ S^\m p^\n +  S^\n p^\m + g^\mn
	\left(mT-(S\c p)\right) \right]\nnn\\
&+&  {\tilde F_1 F_2\over 2m}
\left[ g^\mn\left( T(p\c q)-m(S\c q)\right)
	 + q^\m  \left( m S^\n - T p^\n\right) 
		\right] \nnn \\
&+&  {\tilde F_2 F_1\over 2m}
 \left[ g^\mn\left( T(p\c q)-m(S\c q)\right)
		+ q^\n \left( m S^\m - T p^\m\right) 
		\right] \nnn \\
&+&  {{\tilde F}_2 F_2\over 4m^2} \left[
            g^\mn\left[ q^2(S\c p) + m T q^2 -2(S\c q)(p\c q)
		\right]\right.		\nnn\\
&&\qquad\quad  -q^\m q^\n\left[ (S\c p) + m T\right]
		+ (p\c q)\left( S^\m q^\n + S^\n q^\m\right)\nnn\\
&&\qquad\quad
		\left. + (S\c q)\left( p^\m q^\n + p^\n q^\m\right)
	        - q^2\left(S^\m p^\n + S^\n p^\m\right) 
	 \right] \nnn \\
&+i&  \tilde G_A F_1\epsilon^{\m\n a b} S_a p_b
 + i {\tilde G_A F_2\over 2m}\epsilon^{\m\n a b} q_b 
		\left( mS_a + T p_a \right)
		 \ ,
\eea
with $Z=3(k^2-m^2)^{2}/4$. Analogously, for cc1
\bea\label{trace_cc1}
Z\; \tilde H^{\mu\nu}_{cc1} &=&
\tG_M G_M
\left[ S^\m p^\n +  S^\n p^\m + g^\mn
	\left(mT-(S\c p)\right) \right]\nnn\\
&-&{\tilde F_2 G_M\over2m} (\bk+p)^\m
     (mS^\n + Tp^\n)\nnn\\
&-&{\tG_M F_2\over2m}(\bk+p)^\n
     (mS^\m + Tp^\m)\nnn\\
&+&  {\tilde F_2 F_2\over 4m^2} (\bk+p)^\m(\bk+p)^\n
        \left(mT+(S\c p)\right) \nnn\\
&+i&\ \tG_A G_M\epsilon^{\m\n a b} S_a p_b \ .
\eea
The other tensors used in this 
work can be mapped to 
\equ(trace_cc2) and \equ(trace_cc1)
\begin{itemize}
\item for the electromagnetic hadronic tensors, 
      $\tilde F_{1,2}\rightarrow 
      F_{1,2}$, $\tilde G_A\rightarrow 0$.
\item
for the corresponding FPWIA tensors of
\equ(snt), 
\be\label{reduce}
\left(\matrix{  S^\m \cr 
                 T   \cr 
              }
               \right)
     \rightarrow \ Z \ 
\left(\matrix{  \bk^\m \cr 
                 m    \cr
             }
       \right) \ .
\ee
\end{itemize}
The cc1 [\equ(trace_cc1)] and cc2 [\equ(trace_cc2)] forms 
agree in the ``on-shell'' limit,
consisting of (I) \equ(reduce), and (I)
letting $q^\mu\rightarrow\bar q^\m$, where $\bar q^\m\equiv 
p^\m-\bk^\m$. For example, under \equ(reduce) the $\tilde F_1 F_2$
terms in \equ(trace_cc2) and \equ(trace_cc1) become
\bea\label{ena}
{\rm cc2} &\rightarrow&  \sim \left[
       g^{\m\n} \prod(\bar q,q) - q^\m\bar q^\n \right] \nnn\\
{\rm cc1} &\rightarrow& \sim 
\left[ 2g^{\m\n} (m^2-\bk\cdot p) 
          - (\bk-p)^\m(\bk-p)^\n\right] \nnn\\
 & & = \left[
g^{\m\n} \prod(\bar q,\bar q) - q^\m\bar q^\n\right] \ ,
\eea
which coincide under $q^\mu\rightarrow\bar q^\m$.

\section{}

Here we evaluate the  $(+,-)$ and $(-,+)$ contributions 
to \equ(density), using the relevant spectral function
 [cf.~\equ(rho_traces_off)]. Notice that we have chosen to 
define $\xi^\m$ in terms of the positive energy spinors
$u(\vk,s)$. Thus, $\kp\c\xi =0$ but $\km\c\xi$ = 
$n\c\xi=2\E\xi^0\ne 0$. The spin sums are evaluated using
$\sum_s\xi^\mu_{s,s} = 0$  and 
$\sum_s\sum_r\xi^\mu_{s,r} \xi^\nu_{r,s}$ =
$ 2\left( -g^{\m\n} +  \bk^\m\bk^\n/ m^2\right)$.
Since $\rho_{+-}$ is linear in $\xi_\m$ and 
because of the first of these properties, 
there is no contribution in \equ(gen_snt) 
from the $\delta_{s,r}$ part of the projector in \equ(proj).
Thus, effectively,
\bea\label{pmmp}
\chi^s_+\bar\chi^r_- +\chi^s_-\bar\chi^r_+
&\rightarrow&
{\s{\bk}+m\over 4m}\gamma_5 \s{\xi}{}^*\gamma_5\gamma^0
+\gamma_5\gamma^0 {\s{\bk}+m\over 4m}\gamma_5 \s{\xi}{}^*\nnn\\
&=& \left( \E\s\xi{}^* -(\s{\bk}+m)\xi^{*0} \ri)/2m\ ,
\eea
with $\xi\equiv\xi_{s,r}$ = $\xi_{r,s}^*$. Therefore,
from Eqs.\ (\ref{gen_snt}) and (\ref{rho_traces})
\bea\label{plus_minus}
&&\!\!\!\!\!\!\! \sum_{\rm spins} \rho_{+-;rs} 
  \le(\chi^s_+\bar\chi^r_- +\chi^s_-\bar\chi^r_+\ri)
= {c\over 2m} \sum_{\rm spins} \Bigl\{ 
      \le[ \bk\!\cdot\!\U + m\beta \ri] \xi^0 
 \nnn\\ && \!\!\!\!\!\!\!\times
\le[\xi^{*0}(\s\bk+m) - \s\xi{}^*\E  \ri] 
+\E\prod(\xi,\U) \left[\E\s\xi{}^* \!-\!(\s\bk+m)\xi^{*0}\ri] \Bigr\}
\nnn\\
      &=& -{c\over m^3} \Bigl[
\le(\bk\!\cdot\!\U + m\beta \ri) 
\le[ (\s\bk+m)(m^2\!-\!\E^2) \ri. 
 \nnn\\ && \qquad\qquad - \le.
        \E\le( m^2\gamma^0\!-\!\E\s\bk \ri)\ri] 
+    \E^2 \le[ m^2\s\U -\s\bk\bk\!\cdot\!\U \ri]
\nnn\\ &&\qquad\qquad
        -\E^2(\s\bk+m)\le[ m^2\U^0-\bk\!\cdot\!\U \ri] \Bigr]\nnn\\
&=& -{c\over 2m}  \Bigl(
\le[ -n\!\cdot\!\U + m\beta \ri](\s\bk+m)
        -\le[\bk\!\cdot\!\U + m\beta \ri](\s{n}-m)
 \nnn\\ &&\qquad\qquad
 + \le[ m^2+\prod(\bk,n) \ri] (\s\U-\beta)
 \Bigr) \ ,
\eea
where we have set $c=\fp\fm/12\pi^3\md(k^2\!-\!m^2)^2$, and 
used $\E\g^0=(\thru\bk+\thru n)/2$, 
$\E\U^0$ = $(\bk\!\cdot\!\U + n\!\cdot\!\U)/2$, and
$\E^2$ = $(m^2+\bk\!\cdot\!n)/2$. We finally obtain
\bea\label{plus_minus_2}
&&  \sum_{\rm spins} \le( \rho_{+-;rs} 
             \H^{\m\n}_{+-;rs}+
            \rho_{-+;rs} \H^{\m\n}_{-+;rs} \ri)
  \nnn\\ 
&=& {\fp\over\fm}\ \rho_{--}\H^{\m\n}_{++}
+{\fm\over\fp}\ \rho_{++}\H^{\m\n}_{--}
\nnn\\
&-&  {\fp\fm \le[ m^2+\prod(\bk,n) \ri]\over 12\pi^3\md(k^2\!-\!m^2)^2}
\tr\left\{ {\s\U-\beta\over 2m}
          \bar\Gamma^\mu {\s{p} + m \over 2m} \Gamma^\nu \right\} \ .
\eea
Let us compare \equ(plus_minus_2) with \equ(fact1).
The first two ``mixed'' terms in \equ(plus_minus_2) 
correspond to the $(+,-)$ contribution to the first term in 
the RHS of \equ(fact1), while the last term, which spoils
factorization,  derives from the
second term in the RHS of \equ(fact1), 
as anticipated in \equ(poff).

\end{document}